\documentclass[sn-mathphys-num]{sn-jnl}
\usepackage{graphicx}%
\usepackage{multirow}%
\usepackage{amsthm}%
\usepackage{mathrsfs}%
\usepackage[title]{appendix}%
\usepackage{xcolor}%
\usepackage{textcomp}%
\usepackage{manyfoot}%
\usepackage{booktabs}%
\usepackage{algorithm}%
\usepackage{algorithmicx}%
\usepackage{algpseudocode}%
\usepackage{listings}%
\usepackage{colortbl}

\theoremstyle{thmstyleone}

\theoremstyle{thmstyletwo}%

\theoremstyle{thmstylethree}%

\usepackage[utf8]{inputenc}
\usepackage[english]{babel}
\usepackage{amssymb}
\usepackage{array}
\usepackage{eqparbox}
\usepackage{url}
\usepackage{multirow}
\usepackage{multicol}
\usepackage{graphicx}
\usepackage{booktabs}

\usepackage{textcomp}
\newcommand{\p}{\textup{\texttt{+}}}
\newcommand{\m}{\textup{\texttt{-}}}
\usepackage{tikz}
\usetikzlibrary{shapes,arrows}
\usetikzlibrary{calc}
\usepackage{tikz-3dplot}
\usepackage{pgfplots}
\pgfplotsset{compat=newest}
\usepgfplotslibrary{groupplots}
\usepgfplotslibrary{dateplot}
\usepackage{comment}
\usepackage{amsmath}
\usepackage{comment}
\usepackage{amssymb}
\usepackage{amsthm}
\usepackage{algorithm}
\usepackage{algpseudocode}
\usepackage{url}
\usepackage{nomencl}
\makenomenclature
\usepackage{etoolbox}
\usepackage[official]{eurosym}
\usepackage{booktabs}
\usepackage{multirow}
\usepackage{cancel}
\definecolor{Gray}{gray}{0.9}
\definecolor{LightCyan}{rgb}{0.88,1,1}
\renewcommand\nomgroup[1]{%
  \item[\bfseries
  \ifstrequal{#1}{S}{Indices and Sets}{%
  \ifstrequal{#1}{V}{Variables}{%
  \ifstrequal{#1}{D}{Dual Variables}{%
  \ifstrequal{#1}{B}{Binary Variables}{%
  \ifstrequal{#1}{I}{Incidence matrices}{%
  \ifstrequal{#1}{P}{Parameters}{}}}}}}%
]}
\usepackage{alphalph}

\AtBeginDocument{%
  \AtBeginEnvironment{subequations}{%
    \let\alph\alphalphval%
  }}
\usepackage{pgfplots}

\usepackage{empheq}
\usepackage{tabularx}
\usepackage{soul}
\allowdisplaybreaks
\usepackage{float}
\usepackage{placeins}

\usepackage{setspace}

\definecolor{color10}{rgb}{0.917647058823529,0.917647058823529,0.949019607843137}
\definecolor{color11}{rgb}{0.748039215686275,0.700980392156863,0.812745098039216}
\definecolor{color12}{rgb}{0.933823529411765,0.754411764705882,0.583823529411765}
\definecolor{color13}{rgb}{0.95,0.95,0.65}
\definecolor{color14}{rgb}{0.27843137254902,0.431372549019608,0.631372549019608}
\definecolor{color15}{rgb}{0.824509803921569,0.124509803921569,0.492156862745098}
\definecolor{color16}{rgb}{0.666666666666667,0.372549019607843,0.172549019607843}

\definecolor{color0}{rgb}{0.917647058823529,0.917647058823529,0.949019607843137}
\definecolor{color1}{rgb}{0.298039215686275,0.447058823529412,0.690196078431373}
\definecolor{color2}{rgb}{0.333333333333333,0.658823529411765,0.407843137254902}
\definecolor{color3}{rgb}{0.768627450980392,0.305882352941176,0.32156862745098}
\definecolor{color4}{rgb}{0.505882352941176,0.447058823529412,0.698039215686274}

\definecolor{Gray}{gray}{0.9}
\definecolor{LightCyan}{rgb}{0.88,1,1}

\newcolumntype{Y}{>{\centering\arraybackslash}X}

\newcommand{\uintvar}{u}
\newcommand{\uintvarvec}{\mathbf{u}}
\newcommand{\uintvarvecone}{\mathbf{u}_{1}}
\newcommand{\uintvarvectwo}{\mathbf{u}_{2}}
\newcommand{\vintvar}{v}

\newcommand{\xconvar}{x}
\newcommand{\xconvarvec}{\mathbf{x}}
\newcommand{\yconvar}{y}
\newcommand{\yconvarvec}{\mathbf{y}}

\newcommand{\scontvar}{s}
\newcommand{\tconvar}{t}
\newcommand{\parnx}{{n_\xconvar}}
\newcommand{\parny}{{n_\yconvar}}
\newcommand{\parnu}{{n_u}}
\newcommand{\parnl}{{n_l}}
\newcommand{\indi}{i}
\newcommand{\indj}{j}
\newcommand{\indk}{k}
\newcommand{\indl}{l}

\newcommand{\setX}{\mathbb{X}}
\newcommand{\setY}{\mathbb{Y}}
\newcommand{\setR}{\mathbb{R}}
\newcommand{\setRxone}{\setR^{(\parnx\times1)}}
\newcommand{\setRyoneb}{\{0,1\}^{(\parny\times1)}}
\newcommand{\setRyone}{\setR^{(\parny\times1)}}

\newcommand{\setRnux}{\setR^{(\parnu\times\parnx)}}
\newcommand{\setRnlx}{\setR^{(\parnl\times\parnx)}}
\newcommand{\setRnuy}{\setR^{(\parnu\times\parny)}}
\newcommand{\setRnly}{\setR^{(\parnl\times\parny)}}
\newcommand{\setRnuone}{\setR^{(\parnu\times1)}}
\newcommand{\setRnloneb}{\{0,1\}^{(\parnl\times1)}}
\newcommand{\setRnlone}{\setR^{(\parnl\times1)}}

\newcommand{\bigMvecone}{{M}_1}
\newcommand{\bigMvectwo}{{M}_2}
\newcommand{\bigMvecthree}{{M}_3}
\newcommand{\bigMvecfour}{{M}_4}
\newcommand{\bigMone}{M_1}
\newcommand{\bigMtwo}{M_2}
\newcommand{\bigMthree}{M_3}
\newcommand{\bigMfour}{M_4}
\newcommand{\bigMfive}{M_5}
\newcommand{\bigMsix}{M_6}
\newcommand{\dlambdavec}{\boldsymbol{\lambda}}
\newcommand{\dlambda}{\lambda}
\newcommand{\dmuvec}{\boldsymbol{\mu}}
\newcommand{\dmuvecone}{\boldsymbol{\mu}_{1}}
\newcommand{\dmuvectwo}{\boldsymbol{\mu}_{2}}
\newcommand{\dmuvecthree}{\boldsymbol{\mu}_{3}}
\newcommand{\dmu}{\mu}
\newcommand{\dnuvec}{\boldsymbol{\nu}}
\newcommand{\dnuvecone}{\boldsymbol{\nu}_{1}}
\newcommand{\dnuvectwo}{\boldsymbol{\nu}_{2}}
\newcommand{\dnu}{\nu}
\newcommand{\dgamma}{\gamma}
\newcommand{\dgammavec}{\boldsymbol{\gamma}}
\newcommand{\dgammavecone}{\boldsymbol{\gamma}_{1}}
\newcommand{\dgammavectwo}{\boldsymbol{\gamma}_{2}}
\newcommand{\matA}{A}
\newcommand{\matB}{B}
\newcommand{\matC}{C}
\newcommand{\matD}{D}
\newcommand{\matE}{E}

\newcommand{\matG}{G}
\newcommand{\matH}{H}
\newcommand{\matIone}{\mathbf{e}_{1}}
\newcommand{\matItwo}{\mathbf{e}_{2}}
\newcommand{\matJ}{J}
\newcommand{\matK}{K}
\newcommand{\matL}{L}

\newcommand{\matN}{N}

\newcommand{\matS}{S}

\newcommand{\matV}{V}

\newcommand{\matOmega}{\Omega}
\newcommand{\matPsi}{\Psi}

\newcommand{\funch}{h(\dmuvec,\dnuvec)}

\newcommand{\funcfxy}{f(\xconvarvec,\yconvarvec)}
\newcommand{\funcfxyhat}{\hat{f}(\xconvarvec,\yconvarvec)}

\newcommand{\funcgxy}{g(\xconvarvec,\yconvarvec)}
\newcommand{\funcgxyhat}{\hat{g}(\xconvarvec,\yconvarvec)}

\newcommand{\problplpone}{bard1} 
\newcommand{\problplptwo}{bard2} 
\newcommand{\problplpthree}{candler} 
\newcommand{\problplpfour}{anan} 
\newcommand{\problplpfive}{bard3} 
\newcommand{\problplpsix}{bard4} 
\newcommand{\problplpseven}{clark} 
\newcommand{\problplpeight}{colson} 

\usepackage{scalerel}
\usetikzlibrary{svg.path}
\definecolor{orcidlogocol}{HTML}{A6CE39}
\tikzset{
orcidlogo/.pic={
\fill[orcidlogocol] svg{M256,128c0,70.7-57.3,128-128,128C57.3,256,0,198.7,0,128C0,57.3,57.3,0,128,0C198.7,0,256,57.3,256,128z};
\fill[white] svg{M86.3,186.2H70.9V79.1h15.4v48.4V186.2z}
svg{M108.9,79.1h41.6c39.6,0,57,28.3,57,53.6c0,27.5-21.5,53.6-56.8,53.6h-41.8V79.1z M124.3,172.4h24.5c34.9,0,42.9-26.5,42.9-39.7c0-21.5-13.7-39.7-43.7-39.7h-23.7V172.4z}
svg{M88.7,56.8c0,5.5-4.5,10.1-10.1,10.1c-5.6,0-10.1-4.6-10.1-10.1c0-5.6,4.5-10.1,10.1-10.1C84.2,46.7,88.7,51.3,88.7,56.8z};
}
}
\newcommand\orcidicon[1]{\href{https://orcid.org/#1}{\mbox{\scalerel*{
\begin{tikzpicture}[yscale=-1,transform shape]
\pic{orcidlogo};
\end{tikzpicture}
}{|}}}}
\definecolor{color0}{rgb}{0.917647058823529,0.917647058823529,0.949019607843137}
\definecolor{color1}{rgb}{0.748039215686275,0.700980392156863,0.812745098039216}
\definecolor{color2}{rgb}{0.933823529411765,0.754411764705882,0.583823529411765}
\definecolor{color3}{rgb}{0.95,0.95,0.65}
\definecolor{color4}{rgb}{0.27843137254902,0.431372549019608,0.631372549019608}
\definecolor{color5}{rgb}{0.824509803921569,0.124509803921569,0.492156862745098}
\definecolor{color6}{rgb}{0.666666666666667,0.372549019607843,0.172549019607843}
\usepackage{hyperref}
\usepackage{lineno}
\modulolinenumbers[5]

\begin{document}
\title[Article Title]{A New Computational Approach for Solving Linear Bilevel Programs Based on Parameter-Free Disjunctive Decomposition}

\author*[1]{\fnm{Saeed} \sur{Mohammadi}\orcidicon{0000-0003-1823-9653}}\email{saeedmoh@kth.se}
\author[1]{\fnm{Mohammad~Reza} \sur{Hesamzadeh}\orcidicon{0000-0002-9998-9773}}\email{mrhesa@kth.se}
\author[2]{\fnm{Steven~A.} \sur{Gabriel}}\email{sgabriel@umd.edu}
\author[1]{\fnm{Dina} \sur{Khastieva}\orcidicon{0000-0001-9448-6061}}\email{dinak@kth.se}

\affil*[1]{\orgdiv{School of Electrical Engineering and Computer Science}, \orgname{KTH Royal Institute of Technology}, \orgaddress{\street{Teknikringen 33}, \city{Stockholm}, \postcode{114 28}, \country{Sweden}}}

\affil[2]{\orgdiv{Department of Mechanical Engineering}, \orgname{University of Maryland}, \orgaddress{\street{4298 Campus Dr}, \city{Maryland}, \postcode{207 42}, \state{MD}, \country{USA}}}

\markboth{}{Saeed Mohammadi \MakeLowercase{\textit{et al.}}: A New Computational Approach for Solving Linear Bilevel Programs Based on Parameter-Free Disjunctive Decomposition}

\abstract{Linear bilevel programs (linear BLPs) have been widely used in computational mathematics and optimization in several applications. Single-level reformulation for linear BLPs replaces the lower-level linear program with its Karush-Kuhn-Tucker optimality conditions and linearizes the complementary slackness conditions using the big-M technique. Although the approach is straightforward, it requires finding the big-M whose computation is recently shown to be NP-hard. This paper presents a disjunctive-based decomposition algorithm which does not need finding the big-Ms whereas guaranteeing that obtained solution is optimal. Our experience shows promising performance of our algorithm.}
\maketitle
\section{Introduction}\label{sec:introduction}
The single-level reformulation approach for solving linear bilevel programs (linear BLPs) requires finding disjunctive parameters (big-M parameters) that do not cut off any bilevel-optimal solution.
Heuristic techniques are often used to find the big-M parameters \cite{bard1998practical} although it is well-known that these techniques might fail.
Recent papers \cite{sinha2018review}, \cite{davarikia2018trilevel}, \cite{kovacs2019bilevel}, \cite{xu2019twolevel}, and \cite{zeng2014solving}
\cite{kleinert2019there}, \cite{pineda2019solving}, and \cite{pineda2018efficiently} 
show the computational challenges of finding the correct big-M parameters (and solving BLPs)
which is shown to be strongly NP-hard \cite{hansen1992new}. This in turn makes finding the global optimal solution of linear BLPs quite challenging. Accordingly, we propose a disjunctive-based decomposition (DBD) solution algorithm for solving the linear BLPs which (1) does not require finding big-M parameters and thus (2) it is computationally advantageous. Accordingly, using our proposed DBD algorithm: no heuristic technique is needed to find big-M parameters and the obtained solution is guaranteed to be optimal to the given linear BLP under reasonable assumptions. Table \ref{tab:liter} compares our proposed DBD algorithm with the relevant solution algorithms in the literature.
\begin{table}[h!]
\caption{A comparative study based on the existing literature}\label{tab:liter} 
\centering
\begin{tabularx}{0.85\linewidth}{lYYYYYYYYYY}
\toprule
Solution algorithm &
\cite{kovacs2019bilevel}& 
\cite{xu2019twolevel}& 
\cite{conejo2006decomposition}& 
\cite{davarikia2018trilevel}& 
\cite{zeng2014solving}& 
DBD\\ 
\midrule
Require tuning disjunctive parameters&
No & 
No & 
Yes & 
Yes & 
Yes & 
No \\
\rowcolor{Gray}
Find global optimal solution &
No & 
No & 
No & 
No & 
Yes &
Yes \\
\bottomrule
\end{tabularx}
\end{table}

The rest of this paper is organized as follows. Section \ref{section:DBD} explains the proposed DBD algorithm to solve a general linear BLP. Section \ref{section:illustrative} gives an illustrative example. Section \ref{section:general_cases} presents several general linear BLPs and Section \ref{sec:large_scale_cases} contains large-scale case studies. Section \ref{section:conclusion} concludes the paper.
\section{The Proposed Disjunctive-Based Decomposition (DBD) Algorithm} \label{section:DBD}
The general linear BLP is given in \eqref{blp_general} with matrices $\matA\in\setRxone$, $\matB \text{, }\matG\in\setRyone$, $\matC\in\setRnux$, $\matD\in\setRnuy$, $\matE\in\setRnuone$, $\matH\in\setRnlx$, $\matJ\in\setRnly$,
and $\matN\in\setRnlone$.
\begin{subequations}\label{blp_general}
\begin{align}
&{\underset{\xconvarvec}{\text{Minimize }}}
\funcfxy=\matA^\intercal \xconvarvec + \matB^\intercal \yconvarvec\\
&\text{Subject to: }\matC \xconvarvec + \matD \yconvarvec \le \matE\\
&\qquad\qquad\quad\,\xconvarvec \ge0\\
&\qquad\qquad\quad\,\xconvarvec \in \setX\\
&\qquad\qquad\quad\,\yconvarvec\in\matS_1(\xconvarvec)\\
&\qquad\qquad\quad\,\matS_1(\xconvarvec)\coloneqq{\arg \underset{\yconvarvec}{\min}}\;\funcgxy=\matG^\intercal \yconvarvec\\
&\qquad\qquad\quad\,\text{Subject to: }\matH\xconvarvec+\matJ\yconvarvec \le \matN:\dlambdavec\\
&\qquad\qquad\qquad\qquad\quad\;\;\;\,\yconvarvec\ge0\\
&\qquad\qquad\qquad\qquad\quad\;\;\;\,\yconvarvec\in\setY
\end{align}
\end{subequations}
The sets $\setX\subseteq \setRxone$ and $\setY\subseteq\setRyone$ are polyhedral sets.
Since the lower-level problem ($\substack{\text{\normalsize{min}}\\\yconvarvec}\{\matG^\intercal \yconvarvec| \matH\xconvarvec+\matJ\yconvarvec \le\matN,\yconvarvec\ge0,\yconvarvec\in\setY\}$) is a linear programming (LP) problem, the Karush-Kuhn-Tucker conditions are necessary and sufficient
and the linear BLP \eqref{blp_general} can be reformulated as the following single-level equivalent model SL.
\begin{subequations}
\begin{align}
&\textbf{SL} \coloneqq\underset{\xconvarvec,\yconvarvec,\dlambdavec}{\text{\normalsize{Minimize}}}\; \matA^\intercal \xconvarvec+ \matB^\intercal \yconvarvec\\
&\qquad\;\;\;\text{Subject to: }\matC\xconvarvec+\matD\yconvarvec\le\matE\\
&\qquad\;\;\;\qquad\quad\quad\;\;\,\, 0\le\left(\matN - \matH\xconvarvec - \matJ\yconvarvec\right) \perp \dlambdavec\ge0\\
&\qquad\;\;\;\qquad\quad\quad\;\;\,\, 0\le\left(\matG+ \matJ^\intercal\dlambdavec\right)\perp\yconvarvec\ge0\\
&\qquad\;\;\;\qquad\quad\quad\;\;\,\, \xconvarvec\ge0\\
&\qquad\;\;\;\qquad\quad\quad\;\;\,\, \dlambdavec\in\setRnlone
\end{align}
\end{subequations}
$\dlambdavec\in\setRnlone$ is the vector of Lagrange multipliers for $\matH\xconvarvec+ \matJ\yconvarvec\le \matN$. 
Accordingly, if $({\xconvarvec},{\yconvarvec}$, ${\dlambdavec})$ is a global optimal solution of SL problem, then $({\xconvarvec},{\yconvarvec})$ is a global optimal solution of the original linear BLP \eqref{blp_general}.
We assume that the upper- and lower-level variables $\xconvarvec$ and $\yconvarvec$ are bounded, consistent with practical real-life problems, and that disjunctive parameters $\{M_1, M_2, M_3, M_4\}$ for the lower-level complementarity conditions exist and they are sufficiently large.\\

\textbf{Definition 1:} \emph{A set of positive scalars $\{\bigMone,\bigMtwo,\bigMthree,\bigMfour\}$ is \textbf{LP-correct} if the vector of solutions $\yconvarvec$ to the lower-level problem and vector of solutions $\dlambdavec$ to its dual problem are the same as vectors of solutions $\{\yconvarvec,\dlambdavec\}$ to problem \eqref{blp_kkt_linear_3}-\eqref{blp_kkt_linear_6}.}\\

The complementary slackness condition in SL (designated by $\perp$) have disjunctive property. Therefore it can be linearized with the big-M technique \cite{fortuny1981representation} with LP-correct big-M parameters $\{M_1, M_2, M_3, M_4\}$.
This results an equivalent problem for SL formulated in \eqref{blp_kkt_linear}.
\begin{subequations}\label{blp_kkt_linear}
\begin{align}
& \underset{\;\;\; \xconvarvec,\yconvarvec,\dlambdavec,\uintvarvecone,\uintvarvectwo}{\text{\normalsize{Minimize}}}
\funcfxy=\matA^\intercal \xconvarvec + \matB^\intercal \yconvarvec \label{blp_kkt_linear_0}\\
&\text{Subject to:}\;  \matC \xconvarvec  +  \matD\yconvarvec \le \matE:\dmuvecone \label{blp_kkt_linear_1}\\
&\qquad\qquad\;\;\, \matH\xconvarvec +  \matJ\yconvarvec \le \matN:\dmuvectwo\label{blp_kkt_linear_2}\\
&\qquad\qquad\;\;\, \m\matJ^\intercal\dlambdavec\le\matG :\dmuvecthree\label{blp_kkt_linear_7}\\
&\qquad\qquad\;\;\, \matN -  \matH\xconvarvec  -  \matJ\yconvarvec \le \bigMvecone (\matIone -  \uintvarvecone):\dnuvecone\label{blp_kkt_linear_3} \\
&\qquad\qquad\;\;\, \dlambdavec \le \bigMvectwo \uintvarvecone:\dgammavecone \label{blp_kkt_linear_4}\\
&\qquad\qquad\;\;\, \matG + \matJ^\intercal\dlambdavec \le \bigMvecthree (\matItwo -  \uintvarvectwo):\dnuvectwo\label{blp_kkt_linear_5} \\
&\qquad\qquad\;\;\, \yconvarvec \le \bigMvecfour \uintvarvectwo:\dgammavectwo \label{blp_kkt_linear_6}\\
&\qquad\qquad\;\;\, \xconvarvec\in\setX, \yconvarvec\in\setY \label{blp_kkt_linear_8} \\
&\qquad\qquad\;\;\, \xconvarvec\ge0,\yconvarvec\ge0,\dlambdavec\ge0
\end{align}
\end{subequations}

Where $\uintvarvecone\in\setRnloneb$ and  $\uintvarvectwo\in\setRyoneb$ are vectors of binary variables, the parameters \{$\bigMvecone,\dots,\bigMvecfour$\} are LP-correct, and $\matIone\in\setRnlone$ and $\matItwo\in\setRyone$ are vectors of ones.
In the proposed DBD approach, \eqref{blp_kkt_linear} is replaced by a subproblem (SP) in \eqref{SP} and a master problem (MP) in \eqref{MP} where MP and SP do not depend on any disjunctive parameter (big-M). 
In effect, the role of SP is to find an optimal solution for a given $\uintvarvec$.
The goal of MP is to find an optimal value for a relaxation of \eqref{blp_kkt_linear} with feasibility/optimality cuts added to MP iteratively. The procedure then alternates between SP which provides an upper bound (UB) for the optimal objective value to \eqref{blp_kkt_linear} and MP which provides a lower bound (LB) for that, in order to solve \eqref{blp_kkt_linear}.
From this perspective, our approach has some similarities to the Benders decomposition \cite{benders1962partitioning}.

\begin{figure}[h!]
\centering
  \begin{tikzpicture}[line join = round, line cap = round, auto]
  \pgfmathsetmacro{\factor}{1/sqrt(2)};
  \node[draw,ellipse](start) {\scriptsize Start};
  \node[draw,rectangle, below of=start, node distance=9 mm] (init) {\scriptsize $UB=\infty,LB=\m \infty,\uintvarvec=\hat{\uintvarvec}$, $\matK=\matL=0$};
  \node[draw, fill=color2, rectangle, below of=init, node distance=8 mm] (lp) {\scriptsize Solve SP};
  \node[draw,fill=color2,diamond,text badly centered,below of=lp,node distance=12 mm, aspect=2] (lpinf) {\scriptsize Infeasible?};
  \node[draw, fill=color2, diamond, text badly centered,below of=lpinf,node distance=15 mm, aspect=2] (lpbound) {\scriptsize Bounded?};
  \draw  node[fill, circle, minimum size=0.1pt,right of=lpbound,node distance=1.5cm, inner sep=0pt](lpboundr) {};
  \node[draw, fill=color2, rectangle, below of=lpbound, node distance=13 mm, text badly centered] (lpboundn){\scriptsize $\matL\xleftarrow{}\matL + 1$; Add constraints \eqref{MP_2}};
  \node[draw, fill=color2, rectangle, left of=lpinf, text badly centered, node distance =55 mm, text width=19 em](lpinfy){\scriptsize Add artificial variables to objective function and infeasible constraint(s)};
  \node[draw,fill=color2, rectangle, left of=lpbound, text badly centered, node distance=55 mm, text width=19 em](lpboundy){\scriptsize $\matK\xleftarrow{}\matK + 1$; $\text{Re-index}\,\vintvar_{\indk}$ such that $\matV_{\indk}$ is non-decreasing in $\indk$;\\ 
  Add constraints \eqref{MP_1}; $UB\xleftarrow{}{\min}\{UB,\matV_{\matK}\}$};
  \node[draw,fill=color1, rectangle, below of=lpboundy,node distance = 13 mm, inner sep=5pt, text width=7 em, text badly centered](tsinit){\scriptsize Solve MP};
  \node[draw, fill=color1, diamond, below of=tsinit,node distance =11 mm,text badly centered, aspect=3](tssolis){\scriptsize Infeasible?};
  \node[draw, fill=color1, rectangle, below of=tssolis, node distance=12 mm, text badly centered](tssoly){\scriptsize $LB={\scriptsize \sum} \matV_{\indk} \hat{\vintvar}_\indk$; $\uintvarvec=\hat{\uintvarvec}$};
  \node[draw, fill=color0, diamond, below of=tssoly, node distance=13 mm, aspect=4, text badly centered](tssolis3){\scriptsize \textbf{Step 3: }$UB -  LB>0$};
  \node[draw, rectangle, below of=tssolis3,node distance=13 mm,text badly centered](tssolis3n){\scriptsize Solve SP, Report $\hat{\uintvarvec}$, $\hat{\xconvarvec}$, $\hat{\yconvarvec}$ as optimal solutions};
  \draw  node[fill, circle, minimum size=0.1pt, left of=tssolis3, node distance=38 mm, inner sep=0pt](tssolis3y){\scriptsize };
  \node[draw, fill=color1, rectangle, right of=tssolis,node distance=34 mm](tssolis2n){\scriptsize Report Infeasible};
  \node[draw,fill=color1, ellipse, below of=tssolis2n, node distance=10 mm](stop) {\scriptsize Stop};
  \node[draw, ellipse, below of=tssolis3n, node distance=10 mm](stop2) {\scriptsize Stop};
  \path [draw, -latex'] (start) -- (init);
  \path [draw, -latex'] (init) -- (lp);
  \path [draw, -latex'] (lp) -- (lpinf);
  \path [draw, -latex'] (lpinf) -- node {\scriptsize No} (lpbound);
  \path [draw, -latex'] (lpinf) -- node {\scriptsize Yes} (lpinfy);
  \path [draw, -latex'] (lpinfy) |- (lp);
  \path [draw, -latex'] (lpbound) -- node {\scriptsize No} (lpboundn);
  \path [draw, -latex'] (lpbound) -- node {\scriptsize Yes} (lpboundy);
  \path [draw, -latex'] (lpboundy) -- (tsinit);
  \path [draw, -latex'] (tsinit) -- (tssolis);
  \path [draw, -latex'] (tssolis) -- node {\scriptsize No} (tssoly);
  \path [draw, -latex'] (tssolis) -- node[below] {\scriptsize Yes} (tssolis2n);
  \path [draw, -latex'] (tssolis2n) -- (stop);
  \path [draw, -latex'] (tssoly) -- (tssolis3);
  \path [draw, -latex'] (lpboundn) -- (tsinit);
  \path [draw, -] (tssolis3) -- node {\scriptsize Yes} (tssolis3y);
  \path [draw, -latex'] (tssolis3y) |- (lp);
  \path [draw, -latex'] (tssolis3) -- node {\scriptsize No} (tssolis3n);
  \path [draw, -latex'] (tssolis3n) -- (stop2);
  \draw[very thick,dotted] ($(lpinfy.north west)+(0,0)$) --++ (7.47,0) --++ (0,0.98) --++ (3.95,0) --++ (0,-4.55) -- ($(lpboundn.south west)$) --++ (0,0.95) --++ (-6.5,0) -- ($(lpinfy.north west)+(0,0)$);
  \draw[very thick,dotted] ($(tsinit.north west)+(-2.05,0)$) -- ($(tsinit.north east)$) --++ (0,-1.1) -- ($(tssolis2n.north east)$) --++ (0,-1.75) --++ (-8.2,0) -- ($(tsinit.north west)+(-2.05,0)$);
  \draw ($(lp.north east)+(1.05,-0.22)$) node[rectangle,fill=color2,draw=black,dotted,thick] {\scriptsize \textbf{Step 1}};
  \draw ($(tsinit.north west)+(-1.4,-0.23)$) node[rectangle,fill=color1,draw=black,dotted,thick] {\scriptsize \textbf{Step 2}};
  \end{tikzpicture}
\caption{The proposed DBD algorithm}
\label{fig:DBD}
\end{figure}
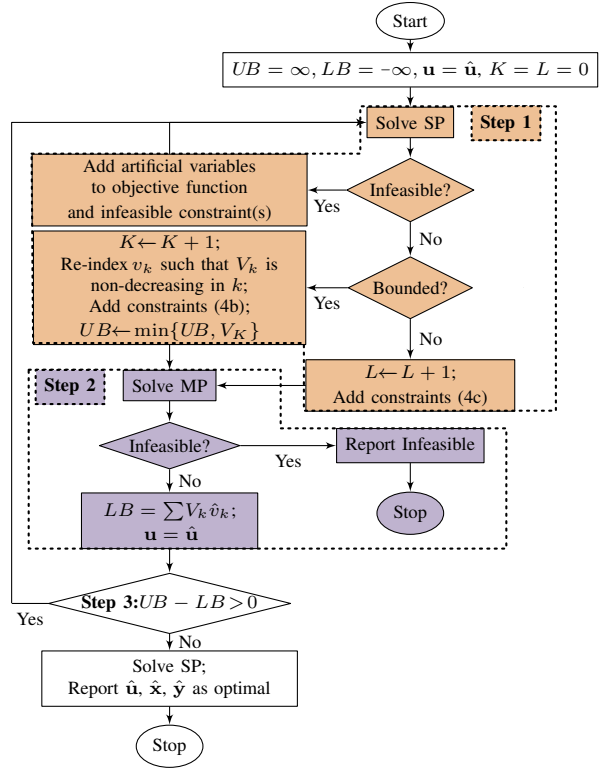

Next, we fix the binary variables $\uintvarvec_1=\hat{\uintvarvec}_1$ and $\uintvarvec_2=\hat{\uintvarvec}_2$ so that \eqref{blp_kkt_linear} is an LP with its dual program as \eqref{DP}.
\begin{subequations}\label{DP}
\begin{align}
& \textbf{DP}
\coloneqq\!\!\!\!\!\!\!\!\!\!\!\!\!\!\!\!\! \underset{\qquad\quad \dmuvec_1, \dmuvec_2, \dnuvec_1, \dnuvec_2, \dgammavec_1, \dgammavec_2}{\text{\normalsize{Maximize}}} 
(\m\dmuvecone^{\intercal}\matE 
\m\dmuvectwo^\intercal\matN 
\m\dmuvecthree^\intercal\matG
\p \dnuvecone^{\intercal} (\matN\m\bigMvecone(\matIone\m\hat{\uintvarvec}_1))  \m
\nonumber \\
& \qquad\qquad\qquad\qquad\qquad\; \dgammavecone^\intercal \bigMvectwo\hat{\uintvarvec}_1  \p \dnuvectwo^{\intercal} (\matG \m \bigMvecthree(\matItwo \m \hat{\uintvarvec}_2))  \m \dgammavectwo^\intercal \bigMvecfour\hat{\uintvarvec}_2)\\
&\qquad\quad \text{Subject to: } 
\m\dmuvecone^\intercal\matC + \left(\dnuvecone^\intercal - \dmuvectwo^\intercal\right)\matH \le \matA^\intercal:\xconvarvec \label{SP_1} 
\\
&\qquad\qquad\qquad\quad\;\;\;
\m \dmuvecone^\intercal\matD + (\dnuvecone^\intercal - \dmuvectwo^\intercal)\matJ  - \dgammavectwo^\intercal\le\matB^\intercal :\yconvarvec \label{SP_2}
\\
\displaybreak[0]
&\qquad\qquad\qquad\quad\;\;\;
\m\dmuvecthree^\intercal\matJ^\intercal - \dgammavecone^\intercal - \dnuvectwo^\intercal\matJ^\intercal\le0:\dlambdavec \label{SP_3} 
\\
\displaybreak[0]
&\qquad\qquad\qquad\quad\;\;\; \dmuvec_1, \dmuvec_2,\dnuvec_1,\dnuvec_2, \dgammavec_1,\dgammavec_2\ge0
\end{align}
\end{subequations}
The DP is equivalent to the SP below as it is shown in Proposition 1. The SP is solved in Step 1 of the proposed DBD algorithm in Fig. \ref{fig:DBD}.
It shows that the dependence on the disjunctive parameters can be dropped under certain circumstances (Proposition 1).
\begin{subequations}\label{SP}
\begin{align}
&\textbf{SP:}\!\!\!\!\!\!\!\!\!\!\!\!\!\! \underset{\qquad\;\;\;\dmuvec_1, \dmuvec_2,\dnuvec_1,\dnuvec_2, \dgammavec_1,\dgammavec_2}{\text{\normalsize{Maximize}}}
 \funch \label{SP_0}\\
\displaybreak[0]
&\;\;\;\;\;\;\, \text{Subject to:} \;\;
\m\dmuvecone^\intercal\matC + \left(\dnuvecone^\intercal - \dmuvectwo^\intercal\right)\matH \le \matA^\intercal:\xconvarvec \label{SP_1} 
\\
\displaybreak[0]
&\qquad\qquad\qquad\;\;  
\m \dmuvecone^\intercal\matD + (\dnuvecone^\intercal - \dmuvectwo^\intercal)\matJ  - \dgammavectwo^\intercal\le\matB^\intercal :\yconvarvec \label{SP_2}
\\
\displaybreak[0]
&\qquad\qquad\qquad\;\;
\m\dmuvecthree^\intercal\matJ^\intercal - \dgammavecone^\intercal - \dnuvectwo^\intercal\matJ^\intercal\le0:\dlambdavec \label{SP_3} 
\\
\displaybreak[0]
&\qquad\qquad\qquad\;\;
\dnuvecone^\intercal (\matIone -  \hat{\uintvarvec}_1) =\dnuvectwo^\intercal (\matItwo - \hat{\uintvarvec}_2) =0\label{SP_4}\\
\displaybreak[0]
&\qquad\qquad\qquad\;\;
\dgammavecone^\intercal \hat{\uintvarvec}_1=\dgammavectwo^\intercal \hat{\uintvarvec}_2 =0\label{SP_5}\\
&\qquad\qquad\qquad\;\; \dmuvec_1, \dmuvec_2,\dnuvec_1,\dnuvec_2, \dgammavec_1,\dgammavec_2\ge0
\end{align}
\end{subequations}\\

\textbf{Proposition 1.} \emph{
Suppose LP-correct parameters $\{\bigMone$, $\bigMtwo$, $\bigMthree$, $\bigMfour\}$ exist.  Furthermore, assume that  these parameters  
are large enough so that: i) $(\matN -  \matH \xconvarvec  -  \matJ \yconvarvec)_\indj < \bigMone$, $\forall \xconvarvec,\yconvarvec,\indj$, and  ii) $(\matG  +  \matJ^\intercal\dlambdavec)_\indj \textless\bigMthree$, $\forall \dlambdavec, \indj$ (element $\indj$ of each vector).
Then, the terms $\dnuvecone^\intercal (\matIone - \hat{\uintvarvec}_1)$,  $\dgammavecone^\intercal\hat{\uintvarvec}_1$,  $\dnuvectwo^\intercal (\matItwo - \hat{\uintvarvec}_2)$, and $\dgammavectwo^\intercal\hat{\uintvarvec}_2$ in the objective function of the DP are zero at an optimal point of \eqref{blp_kkt_linear}. This means that the DP and the SP have the same solution set.}\\

\textit{Proof.} If the binary variable $(\hat{\uintvarvec}_1)_\indj=0$, then row $\indj$ of constraints \eqref{blp_kkt_linear_3} \big($(\matN -  \matH\xconvarvec  -  \matJ\yconvarvec)_\indj \le \bigMvecone$\big) is not binding 
and the corresponding Lagrange multiplier $(\dnuvecone)_\indj=0$. Accordingly the term $(\dnuvecone)_\indj (\matIone - \hat{\uintvarvec}_1)_\indj$ is zero. Otherwise, the binary variable is $(\hat{\uintvarvec}_1)_\indj=1$ and it is clear that $(\dnuvecone)_\indj (\matIone - \hat{\uintvarvec}_1)_\indj$ is zero. Therefore, we have $\dnuvecone^\intercal(\matIone - \hat{\uintvarvec}_1)=0$. The same arguments are valid for the other three terms in {Proposition 1}. Hence, these terms can be removed from the objective function of DP and represented as constraints \eqref{SP_4}-\eqref{SP_5} in the SP.\hfill $\square$\\

In the case of infeasibility of the SP, an alternative always-feasible version is formulated and solved employing artificial variables in the objective function and infeasible constraint(s) \cite{conejo2006decomposition}. 
Observe that SP provides an upper bound to the optimal solution of the mixed-integer LP (MILP) problem in \eqref{blp_kkt_linear} and it does not have any big-M parameters.\\
In Step 2, the MP \eqref{MP} is solved using the optimal values ($\matV_\indk$) of the SP and extreme points $\indk \in\{0,\dots,\matK\}$ and extreme rays $\indl\in\{1,\dots,\matL\}$ found by solving the SP in previous iterations.
If the SP has an extreme point, then index of the binary variables corresponding to $\dnuvec_1>0$ and $\dnuvec_2>0$ are stored in the set $\matOmega_{\indk}$ and the set $\matOmega'_{\indk}$ stores index of the binary variables corresponding to $\dgammavec_1 >0$ and $\dgammavec_2 >0$.
If the SP has an extreme ray, then the set $\matPsi_{\indl}$ stores indices of the binary variables corresponding to $\dnuvec_1>0$ and $\dnuvec_2>0$ and the set $\matPsi'_{\indl}$ stores indices of the binary variable corresponding to $\dgammavec_1 >0$ and $\dgammavec_2 >0$.  $|\matOmega_{\indk}|$, $|\matOmega'_{\indk}|$, $|\matPsi'_{\indl}|$, and $|\matPsi_{\indl}|$ are the cardinality of these sets and \eqref{MP_3} is used to enforce the tree search algorithm.
Lastly, in the objective function \eqref{MP_0}, $\matV_{\indk}$s should be re-ordered such that $\matV_{\indk}$ is non-decreasing in $\indk$.
\begin{subequations} \label{MP}
\begin{align}
&
\textbf{MP: }
\underset{\uintvar_{\indi}, \vintvar_\indk}{\text{\normalsize{Minimize}}}\;\;\;
\sum_{\indk=0}^{\matK} 
\matV_{\indk} \vintvar_\indk \label{MP_0}\\
&\quad\;\;\;\;\;\text{Subject to: } \sum_{\indi \in \matOmega_{\indk}} \uintvar_{\indi}  + \sum_{\indi \in \matOmega'_{\indk}} \uintvar_{\indi} \le |\matOmega_{\indk}| +  |\matOmega'_{\indk}| -  1  + \sum_{\indk' \ge \indk}  \vintvar_{\indk'};~~ \forall \indk\ge1 \label{MP_1}\\
&\qquad\qquad\qquad\quad\;\sum_{\indi\in \matPsi_{\indl}}  \uintvar_{\indi}  +  \sum_{\indi\in \matPsi'_{\indl}} (1 - \uintvar_{\indi}) \le |\matPsi_{\indl}|  +  |\matPsi'_{\indl}| - 1 ;~~\forall \indl\ge1\label{MP_2}\\
&\qquad\qquad\qquad\quad\; \sum_{\indk=0}^{\matK}\;\, \vintvar_\indk = 1\; \\
&\qquad\qquad\qquad\quad\; \uintvar_{\indi}, \vintvar_\indk \in \{0,1\}
\label{MP_3}
\end{align}
\end{subequations}
As it is proved in \cite{bean1992bender}, the MP provides a lower bound for the optimal solution of the MILP problem \eqref{blp_kkt_linear}.
In Step 3, if $UB -  LB=0$ then the DBD algorithm terminates.
Otherwise, the next iteration starts from the new values obtained in Step 2.
Proposed DBD algorithm is detailed in Fig. \ref{fig:DBD} and applied to a simple example in Section \ref{section:illustrative} to clearly explain the process.

\section{Illustrative example} \label{section:illustrative}
In this section, the proposed DBD algorithm for illustrative example is implemented in Julia 1.4.0 \cite{julia2017} and LP problems are solved with GNU linear programming kit (GLPK) 0.12.1 \cite{glpksoftware}.
Detailed operation of our proposed DBD algorithm is shown using the illustrative linear BLP \eqref{L_example} below:
\begin{subequations}\label{L_example}
\begin{align}
&\underset{\;\xconvar\in\setR^{\p}}{\text{\normalsize{Minimize }}}\funcfxy=0.01\xconvar-\yconvar\\
&\text{Subject to: } 0\le\xconvar\le1\\
&\qquad\qquad\;\;\, \yconvar\in\matS_2(\xconvar) \\
&\qquad\qquad\;\;\, \matS_2(\xconvar):=\underset{\yconvar \in\setR^{\p}}{\text{\normalsize{arg min}}}\;\;\; \funcgxy=\yconvar\\
&\qquad\qquad\;\;\, \text{Subject to: } 0.01\yconvar  -  \xconvar\ge\m0.5\\
&\qquad\qquad\qquad\qquad\;\;\;\;\;\, \yconvar  +  \xconvar\ge1
\end{align}
\end{subequations}
$\dlambda_1$ and $\dlambda_2$ are Lagrange variables of constraints $0.01\yconvar  -  \xconvar\ge - 0.5$ and $\yconvar + \xconvar\ge1$, respectively. The single-level MILP reformulation of
\eqref{L_example} is derived in \eqref{L_example_kkt_linear}.
\begin{subequations}\label{L_example_kkt_linear}
\begin{align}
&\!\!\!\!\!\!\!\!\!\!\!\!\!\!\!\!\! \underset{\qquad\;\;\; \xconvar,\yconvar,\dlambda_1,\dlambda_2,\uintvar_{1},\uintvar_{2},\uintvar_{3}}{\text{\normalsize{Minimize}}}\;\;
\funcfxy=0.01\xconvar-\yconvar \label{L_example_kkt_linear_0} \\
&\text{Subject to: }\m\xconvar\ge\m1:\dmu_1 \label{L_example_kkt_linear_1} \\
\displaybreak[0]
&\qquad\qquad\;\;\; 1 - 0.01\dlambda_1 - \dlambda_2\ge0:\dmu_2\label{L_example_kkt_linear_2} \\
\displaybreak[0]
&\qquad\qquad\;\;\; 0.01\yconvar - \xconvar\ge \m0.5:\dmu_3\label{L_example_kkt_linear_3_1}\\
\displaybreak[0]
&\qquad\qquad\;\;\; \yconvar + \xconvar\ge 1:\dmu_4\label{L_example_kkt_linear_3_2}\\
\displaybreak[0]
&\qquad\qquad\;\;\; \xconvar\m0.01\yconvar\ge0.5  - \bigMone (1 - \uintvar_{1}):\dnu_{1}\label{L_example_kkt_linear_5_1} \\
\displaybreak[0]
&\qquad\qquad\;\;\; \m\xconvar  - \yconvar \ge\m1 -  \bigMtwo (1 -  \uintvar_{2}):\dnu_{2}\label{L_example_kkt_linear_5_2} \\
\displaybreak[0]
&\qquad\qquad\;\;\; 0.01\dlambda_1 + \dlambda_2 \ge1 -  \bigMfive (1 -  \uintvar_{3}):\dnu_{3}\label{L_example_kkt_linear_5_3} \\
\displaybreak[0]
&\qquad\qquad\;\;\; \m\dlambda_1\ge\m\bigMthree \uintvar_{1}:\dgamma_{1} \label{L_example_kkt_linear_4_1} \\
\displaybreak[0]
&\qquad\qquad\;\;\; \m\dlambda_2\ge\m\bigMfour \uintvar_{2}:\dgamma_{2}\label{L_example_kkt_linear_4_2} \\
\displaybreak[0]
&\qquad\qquad\;\;\; \m\yconvar\ge\m\bigMsix \uintvar_{3}:\dgamma_{3}\label{L_example_kkt_linear_4_3} \\
&\qquad\qquad\;\;\; \xconvar,\yconvar,\dlambda_1,\dlambda_2\in\setR^{\p} \\
&\qquad\qquad\;\;\; \uintvar_{1},\uintvar_{2},\uintvar_{3}\in\{0,1\}
\end{align}
\end{subequations}
The SP for a given $\hat{\uintvarvec}=[\hat{\uintvar}_{1},\hat{\uintvar}_{2},\hat{\uintvar}_{3}]^\intercal$ is derived in SP-1 below where $\dmuvec=[\dmu_1,\dmu_2,\dmu_3,\dmu_4]^\intercal$, $\dnuvec=[\dnu_{1},\dnu_{2},\dnu_{3}]^\intercal$, and $\dgammavec=[\dgamma_{1},\dgamma_{2},\dgamma_{3}]^\intercal$ are vectors of Lagrange multipliers.
\begin{subequations}\label{ex_SP}
\begin{align}
&\!\!\!\!\!\!\!\!\!\textbf{SP-1: }\underset{\;\;\;\dmuvec,\dnuvec,\dgammavec}{\text{\normalsize{Maximize}}}\;\; \m\dmu_1-\dmu_2-0.5\dmu_3+\dmu_4+0.5\dnu_{1} - \dnu_{2} + \dnu_{3} \label{ex_SP_0}\\
\displaybreak[0]
&\;\;\;\;\;\;\text{Subject to: }\m\dmu_1 - \dmu_3+\dmu_4+\dnu_{1} - \dnu_{2}\le0.01:\xconvar\label{ex_SP_1} \\
\displaybreak[0]
&\qquad\qquad\qquad\;\;\;0.01\dmu_3+\dmu_4-0.01\dnu_{1}-\dnu_{2}-\dgamma_{3}\le\m1:\yconvar \label{ex_SP_2}\\
\displaybreak[0]
&\qquad\qquad\qquad\;\;\;\m0.01\dmu_2+0.01\dnu_{3}-\dgamma_{1}\le0:\dlambda_1 \label{ex_SP_3} \\
\displaybreak[0]
&\qquad\qquad\qquad\;\;\;\m\dmu_2 + \dnu_{3} - \dgamma_{2}\le0\;:\dlambda_2   \label{ex_SP_4} \\
\displaybreak[0]
&\qquad\qquad\qquad\;\;\;\dnu_{1}(1 - \hat{\uintvar}_{1})=\dnu_{2}(1 - \hat{\uintvar}_{2})=\dnu_{3}(1 - \hat{\uintvar}_{3})=0   \label{ex_SP_5}\\
\displaybreak[0]
&\qquad\qquad\qquad\;\;\; \dgamma_{1}\hat{\uintvar}_{1}=\dgamma_{2}\hat{\uintvar}_{2}=\dgamma_{3}\hat{\uintvar}_{3}=0 \label{ex_SP_6}\\
&\qquad\qquad\qquad\;\;\; \dmu_1,\dmu_2,\dmu_3,\dmu_4,\dnu_{1},\dnu_{2},\dnu_{3},\dgamma_{1},\dgamma_{2},\dgamma_{3} \in \setR^{\p}
\end{align}
\end{subequations}
\setlength{\tabcolsep}{2.5 pt}
\begin{table*}[h!]
\caption{SP-1 solutions for all combinations of $\{\hat{\uintvar}_{1},\hat{\uintvar}_{2},\hat{\uintvar}_{3}\}$}
\label{tab:extr_rays} 
\centering
\begin{tabularx}{\linewidth}{llllXllXXXlllXXl}%
\toprule
{\rotatebox[origin=c]{90}{Row}}
& $\hat{\uintvar}_{1}$ & $\hat{\uintvar}_{2}$ & $\hat{\uintvar}_{3}$ & $\hat{\dmu}_{1}$ & $\hat{\dmu}_{2}$ & $\hat{\dmu}_{3}$ & $\hat{\dmu}_4$ & $\hat{\dnu}_{1}$ & $\hat{\dnu}_{2}$ & $\hat{\dnu}_{3}$ & $\hat{\dgamma}_{1}$ & $\hat{\dgamma}_{2}$ & $\hat{\dgamma}_{3}$ & $\hat{\matV}_{\matK}$ & Status
\\ \midrule
1&0 &0 &0&0 &0 &0 &0.01 &0 &0 &0 &0 &0 &1.01 &$\m\infty$ &UNB \\
\rowcolor{Gray}
2& 0 &0 &1&$\times$ &$\times$&$\times$&$\times$&$\times$&$\times$&$\times$&$\times$&$\times$&$\times$ & $\times$ &INF \\ 
3& 0&1&0&0&0&0&0.01&0 &0 &0 &0 &0 &1.01 &$\m\infty$ &UNB \\
\rowcolor{Gray}
4 &0 &1 &1&0  &0 &0 &0 &0 &1 &0 &0 &0 &0 &\m1 & OPT \\
5& 1 &0 &0&0  &0 &0 &0.01&0 &0 &0 &0 &0 &1.01 &$\m\infty$ & UNB \\
\rowcolor{Gray}
6 &1 &0 &1 &99.99 &0 &0 &0 &100 &0 &0 &0 &0 &0 &{\bf \m49.99} &OPT \\
7 &1 &1 &0 &0 &0 &0 &0.01 &0 &0 &0 &0 &0 &1.01 &$\m\infty$ & UNB \\
\rowcolor{Gray}
8 &1 &1 &1&0 &0 &0 &0 &1 &0.99 &0 &0 &0 &0 &\m0.49 & OPT \\
\bottomrule
\end{tabularx}\centering
\\
\begin{tabularx}{\linewidth}{X}
UNB: Unbounded, INF: Infeasible, OPT: Optimal
\end{tabularx}\centering
\\
\end{table*}
\setlength{\tabcolsep}{6pt}

SP-1 is solved for all combinations of $\hat{\uintvar}_{1}$ to $\hat{\uintvar}_{3}$ as shown in Table \ref{tab:extr_rays} only for reference.
The linear BLP problem \eqref{L_example} is solved in four iterations as follows:\\

\textbf{Initialization:} Initialize the proposed DBD algorithm with $UB=10^4$, $LB=\m10^4$, $\hat{\uintvarvec}=[0,0,0]^\intercal$, $\matK=\matL=0$, and $\matV_{0}=\m10^4$.\\

\textbf{Iteration 1:} The SP-1 is solved in Step 1 which is unbounded.
This results in $\matL\xleftarrow{}\matL + 1=1$ and ${\dgamma}_{3}=1.01>0$ from row 1 of Table \ref{tab:extr_rays}. Index sets are $\matPsi_{\matL}=\{\}$ and $\matPsi'_{\matL}=\{3\}$ with cardinalities $|\matPsi_{\matL}|=0$ and $|\matPsi'_{\matL}|=1$.
In Step 2 employing \eqref{MP_2}, we generate the constraint $1\le\uintvar_{3}$ and formulate the MP-1 in \eqref{mp_itr1}.
\begin{subequations} \label{mp_itr1}
\begin{align}
&\!\!\textbf{MP-1: }\underset{\uintvar_{3},\vintvar_{0}}{\text{Minimize}}\;\; \matV_{0} \vintvar_{0}\\
&\qquad\;\;\;\;\text{Subject to: } 1\le\uintvar_{3}\\
&\qquad\qquad\qquad\quad\;\;\; \vintvar_{0}=1\\
&\qquad\qquad\qquad\quad\;\;\; \uintvar_{3},\vintvar_{0}\in\{0,1\}
\end{align}
\end{subequations}
$\uintvar_{1}$ and $\uintvar_{2}$ are free and initialized by zero. Optimal solutions are ${\vintvar}_{0}=1$ and $\hat{\uintvarvec}=[0,0,1]^\intercal$. Step 3 is to go to the next iteration since $LB=\matV_{0}{\vintvar}_{0}=\m10^4$ and $UB -  LB>0$.\\

\textbf{Iteration 2:} The problem SP-1 is infeasible with $\hat{\uintvarvec}=[0,0,1]^\intercal$. Therefore, we add the term ($\m\scontvar - \tconvar$) to the objective function \eqref{ex_SP_0} and change the infeasible constraint \eqref{ex_SP_2} to $0.01\dmu_3+\dmu_4-0.01\dnu_{1}-\dnu_{2}-\dgamma_{3}+\scontvar-\tconvar\le\m1$. In this case, SP-1 becomes feasible but unbounded. This results in $\matL \xleftarrow{}\matL + 1=2$, ${\dnu}_{3}=100>0$,  ${\dgamma}_{1}=1>0$, and ${\dgamma}_{2}=100>0$. Index sets are $\matPsi_{\matL}=\{3\}$ and $\matPsi'_{\matL}=\{1,2\}$ with cardinalities $|\matPsi'_{\matL}|=2$ and $|\matPsi_{\matL}|=1$.
In Step 2 employing \eqref{MP_2}, the constraint $\uintvar_{3}\le\uintvar_{1} + \uintvar_{2}$ is generated in MP-2.
\begin{subequations} \label{mp_itr2}
\begin{align}
&\!\!\textbf{MP-2: }\underset{\uintvar_{1},\uintvar_{2},\uintvar_{3},\vintvar_{0}}{\text{Minimize }}\matV_{0} \vintvar_{0}\\
&\qquad\quad\; \text{Subject to: }  1\le\uintvar_{3}\\
&\qquad\qquad\qquad\qquad  \vintvar_{0}=1\\
&\qquad\qquad\qquad\qquad  \uintvar_{3}\le\uintvar_{1} + \uintvar_{2}\\
&\qquad\qquad\qquad\qquad  \uintvar_{1},\uintvar_{2},\uintvar_{3},\vintvar_{0} \in \{0,1\}
\end{align}
\end{subequations}
Optimal solutions are ${\vintvar}_0=1$ and $\hat{\uintvarvec}=[1,0,1]^\intercal$. Then, Step 3 is to go to the next iteration since $LB=\matV_{0}{\vintvar}_{0}=\m10^4$ and $UB -  LB>0$.\\

\textbf{Iteration 3:} First, we solve the SP-1 with $\hat{\uintvarvec}=[1,0,1]^\intercal$. It is bounded where ${\dnu}_{1}=100>0$, $\matK \xleftarrow{}\matK + 1=1$, and $\matV_{\matK}=\m49.99$. Then, the $UB$ is updated to $UB\xleftarrow{}{\min}\{UB=10^4,\matV_{\matK}=\m49.499\}=\m49.99$. Index sets are $\matOmega_{\matK}=\{1\}$ and $\matOmega'_{\matK}=\{\}$ with cardinalities $|\matOmega_{\matK}|=1$ and $|\matOmega'_{\matK}|=0$.
In Step 2 employing \eqref{MP_1}, the constraint $\uintvar_{1}\le\vintvar_{1}$ is formulated in MP-3.
Indices are not changed since $\matV_{\indk}$s are already non-decreasing in $\indk$ ($\matV_{0}\le\matV_{1}$).
\begin{subequations} \label{mp_itr3}
\begin{align}
&\!\!\textbf{MP-3:}\!\!\! \underset{\;\;\;\;\uintvar_1,\uintvar_2,\uintvar_3,\vintvar_{0},\vintvar_{1}}{\text{Minimize }}
\matV_{0} \vintvar_{0} - 49.99\vintvar_{1}\\
&\qquad\quad\; \text{Subject to: } \vintvar_{0} + \vintvar_{1}=1\\
&\qquad\qquad\qquad\qquad 1\le\uintvar_{3}\le\uintvar_{1} + \uintvar_{2}\\
&\qquad\qquad\qquad\qquad \uintvar_{1}\le\vintvar_{1} \\
&\qquad\qquad\qquad\qquad \uintvar_1,\uintvar_2,\uintvar_3,\vintvar_{0},\vintvar_{1}\in\{0,1\}
\end{align}
\end{subequations}
Optimal solutions are ${\vintvar}_0={1}
$, ${\vintvar}_{1}=0$, and $\hat{\uintvarvec}=[0,1,1]^\intercal$ which result in $LB=\matV_{0}{\vintvar}_{0} + \matV_{1}{\vintvar}_{1}=\m10^4$. We go to the next iteration since $UB -  LB>0$.\\

\textbf{Iteration 4:} The SP-1 is solved with $\hat{\uintvarvec}=[0,1,1]^\intercal$ which is bounded so $\matK\xleftarrow{}\matK + 1=2$, $\matV_{\matK}=\m1$, $\dnu_{2}=1>0$, and $UB\xleftarrow{}{\min}\{U\!B=\m49.99, \matV_{\matK}=\m1\}=\m49.99$. The index sets are $\matOmega_{\matK}=\{2\}$ and $\matOmega'_{\matK}=\{\}$ with cardinalities $|\matOmega_{\matK}|=1$ and $|\matOmega'_{\matK}|=0$. The constraint $\uintvar_{2}\le\vintvar_{2}$ is generated using \eqref{MP_1}. $\matV_{\indk}$s are non-decreasing in $\indk$ which results in MP-4.
\begin{subequations} \label{mp_itr4}
\begin{align}
&\!\!\textbf{MP-4:}\!\!\!\!\!\!\!\! \underset{\;\;\;\;\;\;\;\;\uintvar_1,\uintvar_2,\uintvar_3,\vintvar_{0},\vintvar_{1},\vintvar_{2}}{\text{Minimize }}
\matV_{0} \vintvar_{0} - 49.99\vintvar_{1} - \vintvar_{2}\label{mp_itr4_0}\\
&\qquad\quad\;\text{Subject to: }\vintvar_{0} + \vintvar_{1}=1\\
&\qquad\qquad\qquad\qquad\, 1\le\uintvar_{3}\le\uintvar_{1} + \uintvar_{2}\\
&\qquad\qquad\qquad\qquad\, \uintvar_{1}\le\vintvar_{1} + \vintvar_{2}\\
&\qquad\qquad\qquad\qquad\, \uintvar_{2}\le\vintvar_{2}\label{mp_itr4_1}\\
&\qquad\qquad\qquad\qquad\, \uintvar_1,\uintvar_2,\uintvar_3,\vintvar_{0},\vintvar_{1},\vintvar_{2}\in\{0,1\}
\end{align}
\end{subequations}
Optimal solutions are ${\vintvar}_{0}={\vintvar}_{2}=0$, ${\vintvar}_{1}=1$, and $\hat{\uintvarvec}=[1,0,1]^\intercal$. This results in $LB=\m49.99$ and $UB -  LB=0$. Therefore, there is no need for more iterations and $\uintvarvec=[1,0,1]^\intercal$ is optimal.
The optimal solutions of MILP problem \eqref{L_example_kkt_linear} (or the bilevel-optimal solutions of linear BLP \eqref{L_example}) are $\hat{\uintvarvec}=[1,0,1]^\intercal$, ${\xconvar}=1$, and ${\yconvar}=50$. Note that in iterations 1 to 4, we did not have to find the LP-correct big-M parameters at all.
\section{General case studies} \label{section:general_cases}
Numerical results for eight general case studies are presented in Table \ref{tab:case}, Table \ref{tab:casesol}, Table \ref{tab:casetime}, and Table \ref{tab:caseiter}. Number of variables and constraints in the upper- and lower-level problems are shown in Table \ref{tab:case}. Values of the parameter matrices $\matA$ to $\matN$ are available in the corresponding references.
\begin{table}[h!]
\caption{General case studies}\label{tab:case}
\centering
\begin{tabularx}{0.8\linewidth}{
c!{\vrule width 1pt}
Y!{\vrule width 1pt}
Y!{\vrule width 1pt}
Y!{\vrule width 1pt}
Y!{\vrule width 1pt}
Y!{\vrule width 1pt}
c}
\toprule
Row & 
Problem & 
$\parnx$ & 
$\parny$ & 
$\parnu$ & 
$\parnl$ & 
Reference
\\
\midrule
1 & $\problplpone  $ & 2 & 3 & 2 & 6 & \cite{bard1982explicit} \\ \rowcolor{Gray}
2 & $\problplptwo  $ & 2 & 2 & 2 & 5 & \cite{bard1982explicit} \\
3 & $\problplpthree$ & 2 & 6 & 2 & 6 & \cite{candler1982linear} \\ \rowcolor{Gray}
4 & $\problplpfour $ & 1 & 1 & 1 & 6 & \cite{anandalingam1990solution} \\
5 & $\problplpfive $ & 1 & 2 & 1 & 5 & \cite{bard1991some} \\ \rowcolor{Gray}
6 & $\problplpsix  $ & 1 & 2 & 1 & 5 & \cite{bard1998practical}\\
7 & $\problplpseven$ & 1 & 2 & 1 & 5 & \cite{clark1990bilevel}\\ \rowcolor{Gray}
8 & $\problplpeight$ & 2 & 3 & 1 & 3 & \cite{colson2002bipa}\\
\bottomrule
\end{tabularx}
\centering
\begin{tabularx}{\linewidth}{X}
$\parnx$: Number of variables in the upper-level problem, $\parny$: Number of variables in the lower-level problem, $\parnu$: Number of constraints in the upper-level problem, $\parnl$: Number of constraints in the lower-level problem.
\end{tabularx}
\centering
\end{table}

These problems are solved with both the big-M algorithm and our proposed DBD algorithm. Optimal values of objective functions (${\funcfxyhat}$ and ${\funcgxyhat}$) and optimal solutions of the upper- and lower-level decision variables ($\hat{\xconvarvec}$ and $\hat{\yconvarvec}$) are shown in Table \ref{tab:casesol}. As we can see for these small case studies, both the big-M algorithm and our proposed DBD algorithm achieve the same results. 
\begin{table}[h!]
\caption{Solutions}\label{tab:casesol}
\centering
\begin{tabularx}{0.7\linewidth}{
c!{\vrule width 1pt}
Y!{\vrule width 1pt}
Y!{\vrule width 1pt}
Y!{\vrule width 1pt}
Y!{\vrule width 1pt}
c}
\toprule
Row & 
Problem & 
${\funcfxyhat}$ & 
${\funcgxyhat}$ & 
$\hat{\xconvarvec}$ &
$\hat{\yconvarvec}$ 
\\
\midrule
1 & $\problplpone$ & \m26& 3.2 &[0,0.9] &[0,0.6,0.4] \\ \rowcolor{Gray}
2 & $\problplptwo$ & \m3.25 & \m4 & [2,0] & [1.5,0] \\
3 & $\problplpthree$ & \m29.2 & 3.2 & [0,0.9] & [0,0.6,0.4,0,0,0] \\ \rowcolor{Gray}
4 & $\problplpfour$ & \m49 & 17 & [16] & [11] \\
5 & $\problplpfive$ & \m1 & 0 & [1] & [0,0] \\ \rowcolor{Gray}
6 & $\problplpsix$ & \m2 & \m1 & [0] & [0,1]\\
7 & $\problplpseven$ & \m13 & \m4 & [5]& [4,2]\\ \rowcolor{Gray}
8 & $\problplpeight$ & \m14.6 & 0.3 & [0,0.65] & [0,0.3,0] \\
\bottomrule
\end{tabularx}
\begin{tabularx}{\linewidth}{X}
\end{tabularx}\centering
\end{table}

\begin{figure}[h!]
\centering
\begin{tikzpicture}

\definecolor{color0}{rgb}{0.534313725490196,0.751960784313725,0.534313725490196}
\definecolor{color1}{rgb}{0.748039215686275,0.700980392156863,0.812745098039216}
\definecolor{color2}{rgb}{0.933823529411765,0.754411764705882,0.583823529411765}
\definecolor{color3}{rgb}{0.95,0.95,0.65}
\definecolor{color4}{rgb}{0.27843137254902,0.431372549019608,0.631372549019608}
\definecolor{color5}{rgb}{0.824509803921569,0.124509803921569,0.492156862745098}
\definecolor{color6}{rgb}{0.666666666666667,0.372549019607843,0.172549019607843}

\begin{groupplot}[group style={group size=2 by 2}, width=0.5\linewidth, height=0.4\linewidth]
\nextgroupplot[
tick align=outside,
tick pos=both,
title={Iterations},
x grid style={white!69.01960784313725!black},
xlabel={Iterations},
xmin=0.35, xmax=14.65,
xtick style={color=black},
y dir=reverse,
y grid style={white!69.01960784313725!black},
ylabel={Problem},
ymin=-0.5, ymax=7.5,
ytick style={color=black},
ytick={0,1,2,3,4,5,6,7},
yticklabels={
\problplpone  ,
\problplptwo  ,
\problplpthree,
\problplpfour ,
\problplpfive ,
\problplpsix  ,
\problplpseven,
\problplpeight},
xticklabel style={rotate=90},
xtick={0,2,4,6,8,10,12,14,16},
]
\path [draw=white!23.92156862745098!black, fill=color0, semithick]
(axis cs:8,-0.15)
--(axis cs:8,0.15)
--(axis cs:9,0.15)
--(axis cs:9,-0.15)
--(axis cs:8,-0.15)
--cycle;
\path [draw=white!23.92156862745098!black, fill=color1, semithick]
(axis cs:6,0.85)
--(axis cs:6,1.15)
--(axis cs:11,1.15)
--(axis cs:11,0.85)
--(axis cs:6,0.85)
--cycle;
\path [draw=white!23.92156862745098!black, fill=color2, semithick]
(axis cs:6,1.85)
--(axis cs:6,2.15)
--(axis cs:11,2.15)
--(axis cs:11,1.85)
--(axis cs:6,1.85)
--cycle;
\path [draw=white!23.92156862745098!black, fill=color3, semithick]
(axis cs:5,2.85)
--(axis cs:5,3.15)
--(axis cs:6,3.15)
--(axis cs:6,2.85)
--(axis cs:5,2.85)
--cycle;
\path [draw=white!23.92156862745098!black, fill=color4, semithick]
(axis cs:2,3.85)
--(axis cs:2,4.15)
--(axis cs:3,4.15)
--(axis cs:3,3.85)
--(axis cs:2,3.85)
--cycle;
\path [draw=white!23.92156862745098!black, fill=color5, semithick]
(axis cs:6,4.85)
--(axis cs:6,5.15)
--(axis cs:6,5.15)
--(axis cs:6,4.85)
--(axis cs:6,4.85)
--cycle;
\path [draw=white!23.92156862745098!black, fill=color6, semithick]
(axis cs:10,5.85)
--(axis cs:10,6.15)
--(axis cs:12,6.15)
--(axis cs:12,5.85)
--(axis cs:10,5.85)
--cycle;
\path [draw=white!23.92156862745098!black, fill=white!40.0!black, semithick]
(axis cs:3,6.85)
--(axis cs:3,7.15)
--(axis cs:5,7.15)
--(axis cs:5,6.85)
--(axis cs:3,6.85)
--cycle;
\addplot [semithick, white!23.92156862745098!black]
table {%
8 0
7 0
};
\addplot [semithick, white!23.92156862745098!black]
table {%
9 0
10 0
};
\addplot [semithick, white!23.92156862745098!black]
table {%
7 -0.075
7 0.075
};
\addplot [semithick, white!23.92156862745098!black]
table {%
10 -0.075
10 0.075
};
\addplot [black, mark=diamond*, mark size=2, mark options={solid,fill=white!23.92156862745098!black}, only marks]
table {%
11 0
11 0
11 0
11 0
12 0
11 0
11 0
11 0
12 0
11 0
12 0
11 0
11 0
11 0
11 0
11 0
12 0
11 0
11 0
11 0
11 0
12 0
11 0
};
\addplot [semithick, white!23.92156862745098!black]
table {%
6 1
1 1
};
\addplot [semithick, white!23.92156862745098!black]
table {%
11 1
11 1
};
\addplot [semithick, white!23.92156862745098!black]
table {%
1 0.925
1 1.075
};
\addplot [semithick, white!23.92156862745098!black]
table {%
11 0.925
11 1.075
};
\addplot [semithick, white!23.92156862745098!black]
table {%
6 2
1 2
};
\addplot [semithick, white!23.92156862745098!black]
table {%
11 2
11 2
};
\addplot [semithick, white!23.92156862745098!black]
table {%
1 1.925
1 2.075
};
\addplot [semithick, white!23.92156862745098!black]
table {%
11 1.925
11 2.075
};
\addplot [semithick, white!23.92156862745098!black]
table {%
5 3
4 3
};
\addplot [semithick, white!23.92156862745098!black]
table {%
6 3
7 3
};
\addplot [semithick, white!23.92156862745098!black]
table {%
4 2.925
4 3.075
};
\addplot [semithick, white!23.92156862745098!black]
table {%
7 2.925
7 3.075
};
\addplot [black, mark=diamond*, mark size=2, mark options={solid,fill=white!23.92156862745098!black}, only marks]
table {%
1 3
1 3
8 3
8 3
8 3
8 3
8 3
};
\addplot [semithick, white!23.92156862745098!black]
table {%
2 4
1 4
};
\addplot [semithick, white!23.92156862745098!black]
table {%
3 4
4 4
};
\addplot [semithick, white!23.92156862745098!black]
table {%
1 3.925
1 4.075
};
\addplot [semithick, white!23.92156862745098!black]
table {%
4 3.925
4 4.075
};
\addplot [semithick, white!23.92156862745098!black]
table {%
6 5
6 5
};
\addplot [semithick, white!23.92156862745098!black]
table {%
6 5
6 5
};
\addplot [semithick, white!23.92156862745098!black]
table {%
6 4.925
6 5.075
};
\addplot [semithick, white!23.92156862745098!black]
table {%
6 4.925
6 5.075
};
\addplot [black, mark=diamond*, mark size=2, mark options={solid,fill=white!23.92156862745098!black}, only marks]
table {%
1 5
5 5
1 5
5 5
4 5
1 5
5 5
5 5
5 5
5 5
1 5
5 5
5 5
5 5
5 5
5 5
5 5
5 5
5 5
5 5
1 5
7 5
8 5
8 5
8 5
7 5
8 5
7 5
8 5
7 5
};
\addplot [semithick, white!23.92156862745098!black]
table {%
10 6
7 6
};
\addplot [semithick, white!23.92156862745098!black]
table {%
12 6
14 6
};
\addplot [semithick, white!23.92156862745098!black]
table {%
7 5.925
7 6.075
};
\addplot [semithick, white!23.92156862745098!black]
table {%
14 5.925
14 6.075
};
\addplot [black, mark=diamond*, mark size=2, mark options={solid,fill=white!23.92156862745098!black}, only marks]
table {%
1 6
1 6
1 6
1 6
};
\addplot [semithick, white!23.92156862745098!black]
table {%
3 7
1 7
};
\addplot [semithick, white!23.92156862745098!black]
table {%
5 7
7 7
};
\addplot [semithick, white!23.92156862745098!black]
table {%
1 6.925
1 7.075
};
\addplot [semithick, white!23.92156862745098!black]
table {%
7 6.925
7 7.075
};
\addplot [semithick, white!23.92156862745098!black]
table {%
8 -0.15
8 0.15
};
\addplot [semithick, white!23.92156862745098!black]
table {%
6 0.85
6 1.15
};
\addplot [semithick, white!23.92156862745098!black]
table {%
6 1.85
6 2.15
};
\addplot [semithick, white!23.92156862745098!black]
table {%
5 2.85
5 3.15
};
\addplot [semithick, white!23.92156862745098!black]
table {%
3 3.85
3 4.15
};
\addplot [semithick, white!23.92156862745098!black]
table {%
6 4.85
6 5.15
};
\addplot [semithick, white!23.92156862745098!black]
table {%
12 5.85
12 6.15
};
\addplot [semithick, white!23.92156862745098!black]
table {%
4 6.85
4 7.15
};

\nextgroupplot[
tick align=outside,
tick pos=both,
title={Elapsed time},
x grid style={white!69.01960784313725!black},
xlabel={Elapsed time},
xmin=0.106, xmax=1.514,
xtick style={color=black},
xtick={0,0.2,0.4,0.6,0.8,1,1.2,1.4,1.6},
xticklabels={0.0,0.2,0.4,0.6,0.8,1.0,1.2,1.4,1.6},
y dir=reverse,
y grid style={white!69.01960784313725!black},
ylabel={\scriptsize},
ymin=-0.5, ymax=7.5,
ytick style={color=black},
ytick={0,1,2,3,4,5,6,7},
xticklabel style={rotate=90},
]
\path [draw=white!23.92156862745098!black, fill=color0, semithick]
(axis cs:0.75,-0.15)
--(axis cs:0.75,0.15)
--(axis cs:0.8925,0.15)
--(axis cs:0.8925,-0.15)
--(axis cs:0.75,-0.15)
--cycle;
\path [draw=white!23.92156862745098!black, fill=color1, semithick]
(axis cs:0.61,0.85)
--(axis cs:0.61,1.15)
--(axis cs:1.05,1.15)
--(axis cs:1.05,0.85)
--(axis cs:0.61,0.85)
--cycle;
\path [draw=white!23.92156862745098!black, fill=color2, semithick]
(axis cs:0.64,1.85)
--(axis cs:0.64,2.15)
--(axis cs:1.0725,2.15)
--(axis cs:1.0725,1.85)
--(axis cs:0.64,1.85)
--cycle;
\path [draw=white!23.92156862745098!black, fill=color3, semithick]
(axis cs:0.55,2.85)
--(axis cs:0.55,3.15)
--(axis cs:0.65,3.15)
--(axis cs:0.65,2.85)
--(axis cs:0.55,2.85)
--cycle;
\path [draw=white!23.92156862745098!black, fill=color4, semithick]
(axis cs:0.2975,3.85)
--(axis cs:0.2975,4.15)
--(axis cs:0.39,4.15)
--(axis cs:0.39,3.85)
--(axis cs:0.2975,3.85)
--cycle;
\path [draw=white!23.92156862745098!black, fill=color5, semithick]
(axis cs:0.59,4.85)
--(axis cs:0.59,5.15)
--(axis cs:0.61,5.15)
--(axis cs:0.61,4.85)
--(axis cs:0.59,4.85)
--cycle;
\path [draw=white!23.92156862745098!black, fill=color6, semithick]
(axis cs:0.9925,5.85)
--(axis cs:0.9925,6.15)
--(axis cs:1.17,6.15)
--(axis cs:1.17,5.85)
--(axis cs:0.9925,5.85)
--cycle;
\path [draw=white!23.92156862745098!black, fill=white!40.0!black, semithick]
(axis cs:0.34,6.85)
--(axis cs:0.34,7.15)
--(axis cs:0.51,7.15)
--(axis cs:0.51,6.85)
--(axis cs:0.34,6.85)
--cycle;
\addplot [semithick, white!23.92156862745098!black]
table {%
0.75 0
0.67 0
};
\addplot [semithick, white!23.92156862745098!black]
table {%
0.8925 0
1.1 0
};
\addplot [semithick, white!23.92156862745098!black]
table {%
0.67 -0.075
0.67 0.075
};
\addplot [semithick, white!23.92156862745098!black]
table {%
1.1 -0.075
1.1 0.075
};
\addplot [black, mark=diamond*, mark size=2, mark options={solid,fill=white!23.92156862745098!black}, only marks]
table {%
1.13 0
1.11 0
1.12 0
1.11 0
1.13 0
};
\addplot [semithick, white!23.92156862745098!black]
table {%
0.61 1
0.18 1
};
\addplot [semithick, white!23.92156862745098!black]
table {%
1.05 1
1.1 1
};
\addplot [semithick, white!23.92156862745098!black]
table {%
0.18 0.925
0.18 1.075
};
\addplot [semithick, white!23.92156862745098!black]
table {%
1.1 0.925
1.1 1.075
};
\addplot [semithick, white!23.92156862745098!black]
table {%
0.64 2
0.18 2
};
\addplot [semithick, white!23.92156862745098!black]
table {%
1.0725 2
1.44 2
};
\addplot [semithick, white!23.92156862745098!black]
table {%
0.18 1.925
0.18 2.075
};
\addplot [semithick, white!23.92156862745098!black]
table {%
1.44 1.925
1.44 2.075
};
\addplot [semithick, white!23.92156862745098!black]
table {%
0.55 3
0.46 3
};
\addplot [semithick, white!23.92156862745098!black]
table {%
0.65 3
0.79 3
};
\addplot [semithick, white!23.92156862745098!black]
table {%
0.46 2.925
0.46 3.075
};
\addplot [semithick, white!23.92156862745098!black]
table {%
0.79 2.925
0.79 3.075
};
\addplot [black, mark=diamond*, mark size=2, mark options={solid,fill=white!23.92156862745098!black}, only marks]
table {%
0.19 3
0.18 3
0.85 3
0.86 3
0.84 3
0.9 3
0.91 3
};
\addplot [semithick, white!23.92156862745098!black]
table {%
0.2975 4
0.19 4
};
\addplot [semithick, white!23.92156862745098!black]
table {%
0.39 4
0.51 4
};
\addplot [semithick, white!23.92156862745098!black]
table {%
0.19 3.925
0.19 4.075
};
\addplot [semithick, white!23.92156862745098!black]
table {%
0.51 3.925
0.51 4.075
};
\addplot [semithick, white!23.92156862745098!black]
table {%
0.59 5
0.58 5
};
\addplot [semithick, white!23.92156862745098!black]
table {%
0.61 5
0.64 5
};
\addplot [semithick, white!23.92156862745098!black]
table {%
0.58 4.925
0.58 5.075
};
\addplot [semithick, white!23.92156862745098!black]
table {%
0.64 4.925
0.64 5.075
};
\addplot [black, mark=diamond*, mark size=2, mark options={solid,fill=white!23.92156862745098!black}, only marks]
table {%
0.17 5
0.51 5
0.17 5
0.51 5
0.43 5
0.17 5
0.52 5
0.53 5
0.51 5
0.53 5
0.18 5
0.53 5
0.53 5
0.52 5
0.52 5
0.53 5
0.53 5
0.53 5
0.54 5
0.53 5
0.18 5
0.69 5
0.77 5
0.78 5
0.77 5
0.7 5
0.78 5
0.69 5
0.84 5
0.7 5
};
\addplot [semithick, white!23.92156862745098!black]
table {%
0.9925 6
0.73 6
};
\addplot [semithick, white!23.92156862745098!black]
table {%
1.17 6
1.38 6
};
\addplot [semithick, white!23.92156862745098!black]
table {%
0.73 5.925
0.73 6.075
};
\addplot [semithick, white!23.92156862745098!black]
table {%
1.38 5.925
1.38 6.075
};
\addplot [black, mark=diamond*, mark size=2, mark options={solid,fill=white!23.92156862745098!black}, only marks]
table {%
0.72 6
0.17 6
0.17 6
0.7 6
0.72 6
0.18 6
0.18 6
0.72 6
1.45 6
};
\addplot [semithick, white!23.92156862745098!black]
table {%
0.34 7
0.17 7
};
\addplot [semithick, white!23.92156862745098!black]
table {%
0.51 7
0.72 7
};
\addplot [semithick, white!23.92156862745098!black]
table {%
0.17 6.925
0.17 7.075
};
\addplot [semithick, white!23.92156862745098!black]
table {%
0.72 6.925
0.72 7.075
};
\addplot [black, mark=diamond*, mark size=2, mark options={solid,fill=white!23.92156862745098!black}, only marks]
table {%
0.77 7
};
\addplot [semithick, white!23.92156862745098!black]
table {%
0.79 -0.15
0.79 0.15
};
\addplot [semithick, white!23.92156862745098!black]
table {%
0.62 0.85
0.62 1.15
};
\addplot [semithick, white!23.92156862745098!black]
table {%
0.66 1.85
0.66 2.15
};
\addplot [semithick, white!23.92156862745098!black]
table {%
0.57 2.85
0.57 3.15
};
\addplot [semithick, white!23.92156862745098!black]
table {%
0.37 3.85
0.37 4.15
};
\addplot [semithick, white!23.92156862745098!black]
table {%
0.6 4.85
0.6 5.15
};
\addplot [semithick, white!23.92156862745098!black]
table {%
1.15 5.85
1.15 6.15
};
\addplot [semithick, white!23.92156862745098!black]
table {%
0.43 6.85
0.43 7.15
};
\end{groupplot}

\end{tikzpicture}
\caption{Distribution of solution time and number of iterations for different initial points}
\label{fig:fig_iterations_elapsed_time}
\end{figure}
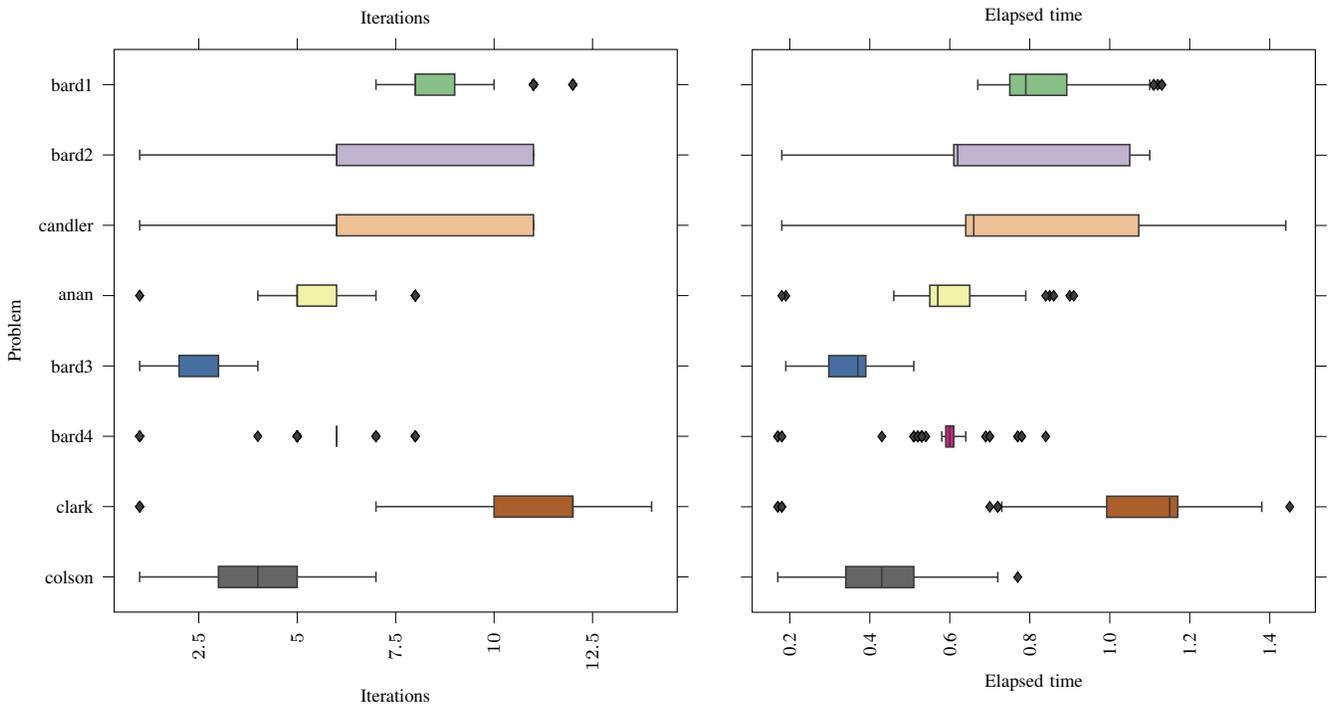

The vector of binary variables $\uintvarvec$ is initialized using an arbitrary binary vector in the proposed DBD algorithm in Fig. \ref{fig:DBD}. To study sensitivity of the proposed DBD algorithm to the initial binary vector $\hat{\uintvarvec}$, eight problems listed in Table \ref{tab:case} are solved with different initial binary vectors. This requires solving each problem up to $2^{(\parny + \parnl)}$ times. For instance, the problem $\problplpthree$ should be solved up to $2^{6 + 6}$=$4,096$ times. To compare results of different problems, 100 random initial binary vectors are selected for each problem. Number of iterations in the proposed DBD algorithm for different initial points $\hat{\uintvarvec}$ is demonstrated in Table \ref{tab:casetime} and Fig. \ref{fig:fig_iterations_elapsed_time}. These calculations are performed on a computer with 2.4 GHz 8-Core processor and memory of 32 GB. Solution time is elapsed real time or wall-time in seconds.
\begin{table}[h!]
\centering
\caption{Solution time (wall-time) and number of iterations for different initial points $\hat{\uintvarvec}$ }\label{tab:casetime}
\begin{tabularx}{0.8\linewidth}{
Y!{\vrule width 1pt}
c!{\vrule width 1pt}
Y!{\vrule width 1pt}
Y!{\vrule width 1pt}
Y!{\vrule width 1pt}
Y!{\vrule width 1pt}
Y!{\vrule width 1pt}
Y!{\vrule width 1pt}
Y!{\vrule width 1pt}
Y}
\toprule
\multirow{2}{*}{\rotatebox[origin=c]{90}{Row}} &
\multirow{2}{*}{\rotatebox[origin=c]{90}{Problem}}
& 
\multicolumn{3}{c}{\qquad Iterations} &&
\multicolumn{3}{c}{\qquad Time (seconds)}
\\ 
&
&\rotatebox[origin=c]{90}{Min} & \rotatebox[origin=c]{90}{Median} &\rotatebox[origin=c]{90}{Mean} &\rotatebox[origin=c]{90}{Max}
&\rotatebox[origin=c]{90}{Min} &\rotatebox[origin=c]{90}{Median} &\rotatebox[origin=c]{90}{Mean} &\rotatebox[origin=c]{90}{Max}\\
\midrule
1
& $\problplpone$
&	7
&	8
&	8.7
&	12
&	0.67
&	0.79
&	0.84
&	1.13
\\ \rowcolor{Gray}
2
& $\problplptwo$
&	1
&	6
&	7.2
&	11
&	0.18
&	0.62
&	0.73
&	1.1
\\
3
& $\problplpthree$
&	7
&	11
&	10.99
&	20
&	0.73
&	1.08
&	1.1
&	1.91
\\ \rowcolor{Gray}
4
& $\problplpfour$
&	1
&	5
&	5.46
&	8
&	0.18
&	0.57
&	0.61
&	0.91
\\
5
& $\problplpfive$
&	1
&	3
&	2.65
&	4
&	0.19
&	0.37
&	0.35
&	0.51
\\ \rowcolor{Gray}
6
& $\problplpsix$ 
&	1
&	6
&	5.72
&	8
&	0.17
&	0.6
&	0.58
&	0.84
\\
7
& $\problplpseven$
&	1
&	12
&	10.78
&	14
&	0.17
&	1.15
&	1.06
&	1.45
\\ \rowcolor{Gray}
8
& $\problplpeight$
&	1
&	4
&	3.79
&	7
&	0.17
&	0.43
&	0.42
&	0.77
\\
\bottomrule
\end{tabularx}
\end{table}

Number of iterations in the big-M algorithm and the proposed DBD algorithm are shown in Table \ref{tab:caseiter} for problems $\problplpone$ to $\problplpeight$. The big-M algorithm requires tuning the big-M parameter for each problem. We observe two issues in tuning the big-M parameters: First, Some problems may become infeasible for some big-M parameters. For example, problems $\problplpone$, $\problplpfour$, $\problplpfive$, and $\problplpsix$ become infeasible with big-M parameters $5$, $10$, $5$, and $5$, respectively. Second, sub-optimal solutions are found with sub-optimal big-M parameters. For instance, employing the big-M algorithm, objective function value for problem $\problplptwo$ with big-M parameter $6$ is $1.75$ instead of optimal objective value which is $\m3.25$. In addition, using large values for the big-M parameters do not guarantee finding optimal solution. For instance, the big-M parameter equal to $5\times10^{5}$ in problem $\problplpone$ leads to objective function value of zero while the optimal objective function value is $\m26$ which is found with M=$10$. These eight problems in Table \ref{tab:caseiter} are small enough that we can tune the big-M parameter and find the optimal solution (reported in Table \ref{tab:caseiter}). Although both big-M and proposed DBD algorithms find the optimal solutions, the proposed DBD algorithm still has the advantage of being big-M free. 
Number of iterations is reported in Table \ref{tab:caseiter} for both algorithms in problems $\problplpone$ to $\problplpeight$. As we can see, the number of iterations in the proposed DBD algorithm is decreased significantly as compared to the one in the big-M algorithm. We have observed a reduction between 60\% and 93\% in number of iterations for all problems except the problem $\problplpseven$ where number of iterations are close (10 and 12 iterations).
\begin{table}[h!]
\caption{Number of iterations in the big-M algorithm and the proposed DBD algorithm}\label{tab:caseiter}
\centering
\begin{tabularx}{0.9\linewidth}{
Y!{\vrule width 1pt}
c!{\vrule width 1pt}
Y!{\vrule width 1pt}
Y!{\vrule width 1pt}
Y!{\vrule width 1pt}
Y!{\vrule width 1pt}
Y!{\vrule width 1pt}
Y!{\vrule width 1pt}
Y!{\vrule width 1pt}
Y}
\toprule
\multirow{3}{*}{\rotatebox[origin=c]{90}{Row}} &
\multirow{2}{*}{\rotatebox[origin=c]{90}{Problem}}
& 
\multicolumn{3}{>{\hsize=\dimexpr3\hsize+3\tabcolsep+\arrayrulewidth\relax}c!{\vrule width 1pt}}{Big-M algorithm}&
\multicolumn{2}{>{\hsize=\dimexpr2\hsize+2\tabcolsep+\arrayrulewidth\relax}c!{\vrule width 1pt}}{Tuned big-M} &
\multicolumn{3}{>{\hsize=\dimexpr3\hsize+3\tabcolsep+\arrayrulewidth\relax}c}{DBD}
\\ 
&
&\rotatebox[origin=c]{90}{big-M} & \rotatebox[origin=c]{90}{Iterations} &\rotatebox[origin=c]{90}{Solution} &\rotatebox[origin=c]{90}{big-M}
&\rotatebox[origin=c]{90}{Iterations} &\rotatebox[origin=c]{90}{Iterations}
&\rotatebox[origin=c]{90}{\%} &\rotatebox[origin=c]{90}{${\funcfxyhat}$}\\
\midrule
1
& $\problplpone$
&	5
&	Inf. 
&	Inf. 
&	10
&	\bf 25
&	\bf 8
&	\bf 68
&	\m26
\\ \rowcolor{Gray}
2
&	$\problplptwo$
&	6
&	13
&	1.75
&	10
&	\bf 15
&	\bf 6
&	\bf 60
&	\m3.25
\\
3
&	$\problplpthree$
&	5
&	28
&	\m23
&	20
&	\bf 29
&	\bf 11
&	\bf 62
&	\m29.2
\\ \rowcolor{Gray}
4
&	$\problplpfour$
&	10
&	Inf.
&	Inf.
&	50
&	\bf 21
&	\bf 5
&	\bf 76
&	\m49
\\
5
&	$\problplpfive$
&	5
&	Inf.
&	Inf.
&	10
&	\bf 14
&	\bf 3
&	\bf 79
&	\m1
\\ \rowcolor{Gray}
6
&	$\problplpsix$ 
&	5
&	Inf.
&	Inf.
&	10
&	\bf 15
&	\bf 6
&	\bf 60
&	\m2
\\
7
&	$\problplpseven$
&	10
&	31
&	\m11
&	100
&	\bf 10
&	\bf 12
&	\bf \m20
&	\m13
\\ \rowcolor{Gray}
8
&	$\problplpeight$
&	10
&	59
&	\m14.6
&	20
&	\bf 57
&	\bf 4
&	\bf 93
&	\m14.6
\\
\bottomrule
\end{tabularx}
\centering
\begin{tabularx}{\linewidth}{X}
Inf.: Infeasible.
\end{tabularx}
\centering
\end{table}

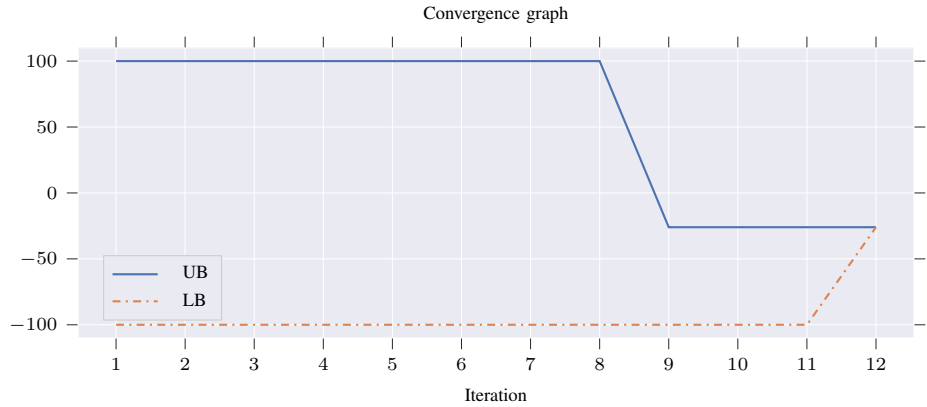
\begin{figure}[h!]
	\centering
\begin{tikzpicture}

\definecolor{color0}{rgb}{0.917647058823529,0.917647058823529,0.949019607843137}
\definecolor{color1}{rgb}{0.298039215686275,0.447058823529412,0.690196078431373}
\definecolor{color2}{rgb}{0.866666666666667,0.517647058823529,0.32156862745098}
\definecolor{color3}{rgb}{0.333333333333333,0.658823529411765,0.407843137254902}
\definecolor{color4}{rgb}{0.768627450980392,0.305882352941176,0.32156862745098}

\begin{axis}[
axis background/.style={fill=color0},
axis line style={white},
legend cell align={left},
legend style={fill opacity=1, draw opacity=1, text opacity=1, at={(0.03,0.06)}, anchor=south west, draw=white!80.0!black, fill=color0},
tick align=outside,
tick pos=both,
title={\scriptsize Convergence graph},
x grid style={white},
xlabel={\scriptsize Iteration},
xmajorgrids,
xmin=0.45, xmax=12.55,
xtick style={color=white!15.0!black},
y grid style={white},
ylabel={\scriptsize },
ymajorgrids,
ymin=-110, ymax=110,
ytick style={color=white!15.0!black},
width=0.7\linewidth, height=0.3\linewidth,
xticklabel style={font=\scriptsize},
yticklabel style={font=\scriptsize},
xtick ={1,2,3,4,5,6,7,8,9,10,11,12},
]
\addplot [thick, color1]
table {%
1 100
2 100
3 100
4 100
5 100
6 100
7 100
8 100
9 -26
10 -26
11 -26
12 -26
};
\addlegendentry{\scriptsize UB}
\addplot [thick, color2, dash dot]
table {%
1 -100
2 -100
3 -100
4 -100
5 -100
6 -100
7 -100
8 -100
9 -100
10 -100
11 -100
12 -26
};
\addlegendentry{\scriptsize LB}
\end{axis}

\end{tikzpicture}
\centering
\caption{The convergence graph of the proposed DBD algorithm}
\label{fig:fig_convergence}
\end{figure}

The convergence graph of the proposed DBD algorithm for problem $\problplpone$ is shown in Fig. \ref{fig:fig_convergence}. The UB and LB values are shown for iterations 1 to 12. The UB and LB are initialized with $100$ and $\m100$, respectively. UB is decreased to $\m26$ at iteration 9 and LB is increased to $\m26$ at iteration 12 where the proposed DBD algorithm stops.

\section{Large-scale case studies}\label{sec:large_scale_cases}
Performance of the proposed approach in large-scale case studies is investigated in this section. First, linear BLPs are defined with randome numbers. Then, these problems are solved both with the big-M algorithm \cite{fortuny1981representation} as well as our proposed DBD algorithm.
\setlength{\tabcolsep}{4pt}
\begin{table}[h!]
\caption{The large-scale case studies}\label{tab:largecase}
\centering
\begin{tabularx}{1.0\linewidth}{
Y!{\vrule width 1pt}
c!{\vrule width 1pt}
c!{\vrule width 1pt}
c!{\vrule width 1pt}
c!{\vrule width 1pt}
Y!{\vrule width 1pt}
Y!{\vrule width 1pt}
Y!{\vrule width 1pt}
Y!{\vrule width 1pt}
Y!{\vrule width 1pt}
Y
}
\toprule
\multirow{2}{*}{Problem} & \multirow{2}{*}{$\parnx$} & \multirow{2}{*}{$\parny$}& \multirow{2}{*}{$\parnu$} & \multirow{2}{*}{$\parnl$}  &	
\multicolumn{3}{>{\hsize=\dimexpr3\hsize+3\tabcolsep+\arrayrulewidth\relax}c!{\vrule width 1pt}}{Average number of Iterations}	
&	
\multicolumn{3}{>{\hsize=\dimexpr4\hsize+4\tabcolsep+\arrayrulewidth\relax}c}{Average optimal value}	
\\
& & & &	& big-M & DBD & \%	&	big-M	&	DBD	&	
{\%} 
\\
\midrule
\textit{Case 1}	&	25 & 25	& 15 &	15	&	840	&	287	&	65.9	&	2.4912	&	2.6007	&
{4.39}
	\\	\rowcolor{Gray}
\textit{Case 2}	&	50 & 50	&	25 & 25	&	8,947	&	371	&	95.85	&	2.4619	&	2.5022	&	
{1.64}
	\\
\textit{Case 3}	&	75& 75	& 50 &	50	&	457,002	&	457	&	99.9	&	4.6533	&	4.7974	&
{3.09}
	\\	\rowcolor{Gray}
\textit{Case 4}	&	100 & 100	&	75 & 75	&	714,711	&	688	&	99.9	&	11.6428	&	11.6735	&
{0.26}
	\\
\bottomrule
\end{tabularx}
\end{table}
\setlength{\tabcolsep}{6pt}

\subsection{Case-study parameters}
Four large-scale case studies (Case1 to Case4) are defined with $\parnx$ = $\parny$ = 25, 50, 75, and 100, respectively to study the effect of problem size on performance of the big-M and the DBD algorithms. Number of constraints are $\parnu$ = $\parnl$ = 15, 25, 50, and 75, respectively. The proposed DBD algorithm does not have any disjunctive parameter (such as the big-M). The big-M parameter is set to 100 in all case studies for the big-M algorithm. The parameter matrices are defined with random numbers as shown in \eqref{cases_param}. Normal(\textit{Mean}, \textit{Std.}) is a function which gives a random number from normal distribution with the given mean (\textit{Mean}) and standard deviation (\textit{Std.}). Uniform(\textit{Min}, \textit{Max}) is a function which gives a random number from the uniform distribution with the given minimum and maximum values (\textit{Min} and \textit{Max}).
\begin{subequations}\label{cases_param}
\begin{align}
&
\matA = [\text{Uniform}( 0,1)]^{(\parnx \times 1     )  };\;
\matB = [\text{Uniform}( 0,1)]^{(\parny \times 1     )  };\; 
\\
&
\matC = [\text{Normal}( 0,1)]^{(\parnx \times \parnu)  };\;
\matD = [0]^{(\parny \times \parnu)  };\;
\\
&
\matE = [\text{Normal}( 0,1)]^{(\parnu \times 1     )  };\;
\matG = [\text{Uniform}( 0,1)]^{(\parny \times 1     )  };\;
\\
&
\matH = [\text{Normal}( 0,1)]^{(\parnx \times \parnl)  };\;
\matJ = [\text{Normal}( 0,1)]^{(\parny \times \parnl)  };\;
\\
&
\matN = [\text{Normal}( 0,1)]^{(\parnl \times 1     )  };\;
\end{align}
\end{subequations}

The matrix $\matD$ is set to zero to avoid infeasible problems. Uniform distribution with values between zero and one are employed in $\matA$, $\matB$, and $\matG$ in the objective functions to avoid unbounded problems. Other parameters are determined from a normal distribution with mean and standard deviation equal to zero and one, respectively.
Since all parameters ($\matA$ to $\matN$) are obtained using random numbers, one hundred problems are defined for each case to mitigate effect of the random values. This allows us to focus mainly on the performance of our proposed DBD algorithm for different problem sizes.

All problems in Section \ref{section:general_cases} and Section \ref{sec:large_scale_cases} are modeled in GAMS 25.1.3 \cite{gamssoftware} and solved with CPLEX 12.8.0.0 solver \cite{ilog2018cplex}. The optimality gap for CPLEX solver (optCR option) is set to zero. The branch and cut algorithm of the CPLEX solver is used for MIP problems. Pre-solving option of the CPLEX solver are disabled to have a fair comparison between results of the CPLEX solver and the proposed DBD algorithm \cite{ilog2018cplex}. All simulations for the large-scale case studies are performed using Tegner processing systems available at KTH PDC Center for High Performance Computing (HPC).

Average number of iterations and average of optimal objective values for \textit{Case 1} to \textit{Case 4} are shown in Table \ref{tab:largecase}. This tables demonstrates significant reduction in number of iterations in our proposed DBD algorithm. For \textit{Case 1}, the number of iterations in the DBD algorithm is decreased by 65\% compared with the big-M algorithm. This is while, the big-M algorithm finds a sub-optimal solution with optimality gap of 4.39\% (the DBD algorithm finds the optimal solution with zero optimality gaps in all cases). For \textit{Case 2}, the DBD algorithm finds the optimal solution in 371 iterations as compared to the 8947 iterations from the big-M algorithm (a reduction of 95.85\%). Also, the big-M algorithm solution has an optimality gap of 1.64\%. For \textit{Case 3} and \textit{Case 4}, the reduction in the number of iterations is significant (99.9\%). The big-M algorithm finds sub-optimal solutions with optimality gaps of 3.09\% and 0.26\%, respectively. As mentioned before, the proposed DBD algorithm finds the optimal solutions for both \textit{Case 3} and \textit{Case 4} with zero duality gap.  

\section{Conclusion}\label{section:conclusion}
This letter proposes a disjunctive-based decomposition (DBD) approach to solve linear bilevel programs (linear BLPs). The proposed DBD algorithm does not need selecting LP-correct big-M parameters which is a hard (if not impossible) process. Accordingly, our obtained solutions are guaranteed to be bilevel optimal without checking the LP-correctness of the big-M parameters as long as the conditions of the Proposition 1 are met. Effectiveness of our proposed DBD algorithm is shown through 12 different case studies. All these studies confirm promising performance of our proposed DBD algorithm. Also, different applications of the proposed DBD algorithm are reported in \cite{dina2021optimal} and \cite{khastieva2020transmission} addressing electricity network investment problem and in  \cite{moiseeva2018} addressing optimal bidding problem of a hydropower generator. 

\section*{Declarations}
This work was financially supported by the Swedish Energy Agency (Energimyndigheten) under Grant 3233. The required computation is performed by computing resources from the Swedish National Infrastructure for Computing (SNIC) at PDC center for high performance computing at KTH Royal Institute of Technology which was supported by the Swedish Research Council under Grant 2018-05973. (Corresponding author: Saeed Mohammadi.)

\section*{References}
\bibliography{bib}
\end{document}